\documentclass[a4paper,reqno,12pt]{JHEP}
\usepackage{amsmath,amsfonts,amssymb,amsthm,amstext,eucal,bbold}
\usepackage{array}
\usepackage[T1]{fontenc}
\usepackage[latin1]{inputenc}
\DeclareMathOperator{\tr}{Tr}
\newcommand{\half}{\tfrac12}
\newcommand{\Neq}[1]{\mathcal{N}{=}#1}
\DeclareMathOperator{\SO}{SO}
\DeclareMathOperator{\ISO}{ISO}
\DeclareMathOperator{\SU}{SU}
\DeclareMathOperator{\U}{U}
\DeclareMathOperator{\SL}{SL}
\DeclareMathOperator{\Spin}{Spin}
\DeclareMathOperator{\CSpin}{CSpin}
\DeclareMathOperator{\dS}{dS}
\DeclareMathOperator{\AdS}{AdS}
\newcommand{\RR}{\mathbb{R}}
\newcommand{\HH}{\mathbb{H}}

\newcommand{\ZZ}{\mathbb{Z}}
\newcommand{\PP}{\mathbb{P}}
\newcommand{\KK}{\mathbb{K}}
\newcommand{\MM}{\mathbb{M}}
\newcommand{\DD}{\mathbb{D}}

\newcommand{\drep}[1]{\underline{\boldsymbol{#1}}}
\newcommand{\sD}{\mathsf{D}}
\newcommand{\sK}{\mathsf{K}}
\newcommand{\sM}{\mathsf{M}}
\newcommand{\sP}{\mathsf{P}}
\newcommand{\sQ}{\mathsf{Q}}
\newcommand{\sR}{\mathsf{R}}
\newcommand{\sS}{\mathsf{S}}
\newcommand{\eD}{\mathcal{D}}

\renewcommand{\d}{\partial}
\newcommand{\sd}{{\mathsf{d}}}
\newcommand{\ssd}{{\mathsf{d}}}
\newcommand{\ssk}{{\mathsf{k}}}
\newcommand{\sm}{{\mathsf{m}}}
\newcommand{\ssm}{{\mathsf{m}}}
\newcommand{\ssp}{{\mathsf{p}}}

\title{Conformal topological Yang--Mills theory and de Sitter holography}

\author{Paul de Medeiros, Christopher Hull, Bill Spence\\
  Department of Physics, Queen Mary, University
  of London, London E1 4NS, England\\
  \email{\{p.demedeiros,c.m.hull,w.j.spence\}@qmul.ac.uk}}

\author{José Figueroa-O'Farrill\\Department of Mathematics and
  Statistics, The University of Edinburgh, Edinburgh EH9 3JZ,
  Scotland\\
  \email{j.m.figueroa@ed.ac.uk}}

\preprint{EMPG-01-04, QMUL-PH-01-16, USYD-01-04}

\abstract{
A new topological conformal field
 theory   in four Euclidean 
dimensions is constructed from $\Neq4$
  super Yang--Mills theory
  by twisting the whole of the conformal group with the whole of the
R-symmetry group, resulting in a theory that is conformally invariant and 
has two conformally invariant BRST operators.
A curved space generalisation is found on any Riemannian 4-fold. This 
formulation
 has local  Weyl invariance and two Weyl-invariant
 BRST symmetries, with an action and energy-momentum
tensor that are BRST-exact.
This theory is expected to have a holographic dual in 5-dimensional de Sitter
space.}


\begin{document}

\section{Introduction}

In \cite{Hull-Timelike} it was proposed that theories in D-dimensional
de Sitter space could have a holographic dual which is a conformal
field theory in D-1 dimensional Euclidean space. In particular, in
\cite{Hull-Timelike, HullKhuri} supersymmetric theories were
identified for which an argument analogous to that of \cite{Maldacena}
for anti-de Sitter holography could be made.  These included a five
dimensional de Sitter vacuum arising from a solution of the type
$\text{IIB}^*$ string theory. This solution preserves 32
supersymmetries and also arises as a solution of a five dimensional
gauged supergravity. The proposed dual theory is the $\Neq4$
superconformal Yang-Mills theory in four Euclidean dimensions.
Unfortunately, both the D=5 supergravity and the D=4
super-Yang-Mills theories are non-standard in that they have some
fields with kinetic terms of the wrong sign.  However, it was pointed
out in \cite{Hull-Timelike} that the super-Yang-Mills theory that
arises in this way is precisely the one that can be twisted to obtain
a topological field theory.  This should correspond to a twisting of
the five dimensional supergravity theory, so that the 't~Hooft limit
of the topological field theory should have a dual description as a
topological supergravity theory in five dimensional de Sitter space.

In four Euclidean dimensions, the Lorentz group is $\Spin(4) \cong
\SU(2) \times \SU(2)$. In the usual twistings, $\Spin(4) $ or an
$\SU(2)$ subgroup thereof is identified with a subgroup of the
R-symmetry group. In the $\Neq4$ theories, this necessarily breaks the
R-symmetry group.  It was pointed out in \cite{Hull-Timelike} that the
R-symmetry group is in fact the same as the conformal group in such
theories, so there is the possibility of twisting the whole of the
R-symmetry group with the conformal group, and this seems the most
natural twisting to use in the holographic context.  The purpose of
this paper is to further develop these proposals, and in particular to
construct a new topological conformal field theory in which the
conformal group is twisted with the whole of the R-symmetry group.

It will be useful to begin by reviewing two ways of viewing
topological field theories.  Consider first the twisting of $\Neq2$
supersymmetric Yang--Mills to give a topological theory whose
observables are the Donaldson invariants \cite{WittenTFT}.  The
$\Neq2$ supersymmetric Yang--Mills theory in 3+1 dimensions can be
obtained by reducing D=6 $\Neq2$ supersymmetric Yang--Mills theory on
a spatial 2-torus.  The D=6 theory has an $\SU(2)$ R-symmetry and the
reduction gives a theory with $\U(2)$ R-symmetry, the extra $\SO(2)$
being related to rotations in the 5-6 plane. (This $\SO(2)$ would be
broken if one were to keep the massive Kaluza-Klein modes.)  The
bosonic sector is Lie-algebra valued and consists of a vector field
and two scalars, which transform as a 2-vector under the $\SO(2)$.
The Wick rotated version of this theory is usually taken to be a
theory in four Euclidean dimensions with $\U(2)$ R-symmetry and the
same bosonic sector. There are the usual subtleties as to how one
treats the fermions, but however one proceeds, the resulting theory in
four Euclidean dimensions has no conventional supersymmetry.

However, there is a simple way of obtaining a supersymmetric theory in
four Euclidean dimensions. One can simply start with the D=6 theory
and reduce on one space and one time dimension, and the resulting
theory in four Euclidean dimensions will automatically be invariant
under $\Neq2$ supersymmetry \cite{BTESYM,AFOS}.  The R-symmetry of
this theory is $\SU(2)\times \SO(1,1)$ and one of the scalar fields
(the one arising from the time component of the D=6 vector field) has
a kinetic term of the wrong sign; this sign is necessary for the
non-compact R-symmetry and for invariance under Euclidean
supersymmetry.  Following \cite{Hull-Timelike}, it will be convenient
to refer to this as a \emph{Euclidean field theory} and to the
Wick-rotated one as a \emph{Euclideanised field theory}.

The twisting of \cite{WittenTFT} was originally formulated in terms of
the Euclideanised theory, with symmetry $\Spin(4) \times \U(2) \cong
\SU(2) \times \SU(2) \times \SU(2) \times \SO(2)$.  An $\SU(2)$
subgroup of the $\Spin(4)$ Lorentz symmetry is twisted with the
$\SU(2)$ subgroup of the $U(2)$ R-symmetry, so that one of the
supercharges becomes a scalar BRST charge.  Equivalently, different
twistings correspond to regarding different embeddings of $\Spin(4)
\cong \SU(2) \times \SU(2)$ in $\Spin(4) \times \SU(2) \cong \SU(2)
\times \SU(2) \times \SU(2)$ as the Lorentz symmetry group.  However,
in this approach, there are some subtleties in finding the action
invariant under the twisted supersymmetry, and in particular, the sign
of the kinetic term of one of the two original scalars must be
changed, so that the twisted theory has an $\SO(1,1)$ symmetry instead
of the original $\SO(2)$ \cite{WittenTFT}; this $\SO(1,1)$ symmetry is
the ghost-number symmetry, with the charge of a field being its
ghost-number.  In performing the functional integral, the problematic
scalar is usually analytically continued $\phi \to i \phi$. The
negative-norm states corresponding to this field are not in the BRST
cohomology and in this sense are not physical.

However, this topological field theory with $\SO(1,1)$ ghost-number
symmetry can be constructed directly by twisting the Euclidean $\Neq2$
supersymmetric Yang--Mills theory \cite{BTESYM,AFOS}.  In this case
the twisting of an $\SU(2)$ subgroup of the $\Spin (4)$ with the
$\SU(2)$ subgroup of the $SU(2)\times SO(1,1)$ R-symmetry is
straightforward and automatically gives a theory invariant under
twisted supersymmetry, and the $\SO(1,1)$ ghost number symmetry is
inherited directly from the Euclidean theory.  For the Euclideanised
theory, the extra sign changes needed to obtain twisted supersymmetry
corresponded in the untwisted theory to changing the
non-supersymmetric Euclideanised theory with $\SU(2)\times \SO(2)$
R-symmetry to the supersymmetric Euclidean one with $\SU(2)\times
\SO(1,1)$ R-symmetry.  The construction from twisting the Euclidean
theory is clearly the more economical and straightforward.

The situation is similar for the $\Neq4$ theories.  Reducing
supersymmetric Yang--Mills from $9+1$ to $4$ dimensions can give
either a Lorentzian theory in $3+1$ dimensions with $\Spin(6)$
R-symmetry or to a theory in four Euclidean dimensions with
$\Spin(5,1)$ R-symmetry.  Both theories have a vector and six scalars,
which transform as a 6-vector under the R-symmetry, and both are
invariant under $\Neq4$ supersymmetry.  Wick rotating the Lorentzian
theory gives the Euclideanised theory with $\Spin(6)$ R-symmetry and
no conventional supersymmetry.  Both the Euclidean and the
Euclideanised theories have 6 scalars, but in the Euclidean theory,
one of the scalars has a kinetic term of the wrong sign.

The corresponding topological field theories were constructed from the
twisting of the $\Spin(4) \cong \SU(2)_L \times \SU(2)_R$ Lorentz
symmetry of the Euclideanised theory with a $\Spin(4) \cong
\SU(2)_1\times \SU(2)_2$ subgroup of the R-symmetry, embedded as
\begin{equation}
  \label{eq:asda}
\Spin(4)\times \Spin(2)\subset \Spin(6).
\end{equation}
There are three inequivalent topological twistings
\cite{Yamron,VafaWitten,Marcus,BTNT2too}.  In the half-twisted model,
one twists $\SU(2)_L$ by $\SU(2)_1$, in the B-model or diagonal
twisting one in addition twists $\SU(2)_R$ by $\SU(2)_2$, while in the
A-model one twists $\SU(2)_L$ by the diagonal subgroup $\SU(2)_D$ of
$\SU(2)_1\times \SU(2)_2$.  In constructing the invariant twisted
action, it is necessary to change some signs and in particular the
sign of the kinetic term of one of the scalars, so that the $\SO(2)$
R-symmetry in \eqref{eq:asda} becomes an $\SO(1,1)$, which is the
ghost-number symmetry.

Again, the twistings can be constructed more directly from the
Euclidean theory, which automatically generates a theory with twisted
supersymmetry without needing to add further sign changes by hand.  In
this case, the Lorentz symmetry is twisted with the $\Spin(4)$
subgroup embedded in the R-symmetry group as
\begin{equation}
  \label{eq:asda2}
  \Spin(4)\times \Spin(1,1)\subset \Spin(5,1)
\end{equation}
in one of the three inequivalent ways described above.

There are then three related Yang-Mills theories in four dimensions,
the Lorentzian one, the Euclidean one and the Euclideanised one.  They
have Lorentz and R-symmetries given respectively by
\begin{equation}
  \label{eq:fgfd}
  \Spin(3,1)\times \Spin(6), \qquad \Spin(4)\times \Spin(5,1), \qquad
  \Spin(4)\times \Spin(6). 
\end{equation}
The first two have 16 supersymmetries, while the Euclideanised one has
none.  Each of these theories is in fact conformally invariant, so
that the Lorentz group is enlarged to the conformal group, and the
three theories have bosonic symmetries given by
\begin{equation}
  \label{eq:fgfd2}
  \Spin(4,2)\times \Spin(6), \qquad \Spin(5,1)\times \Spin(5,1), \qquad
  \Spin(5,1)\times \Spin(6), 
\end{equation}
respectively. The Lorentzian and Euclidean theories are in fact
superconformally invariant, with conformal supergroups $\SU(2,2|4)$
and $\SU^*(4|4)$ respectively.

The three twistings described above all explicitly break the
R-symmetry. However, when viewed as a twisting of the Euclidean
theory, there is in addition a fourth possible twisting
\cite{Hull-Timelike}.  The Euclidean theory has symmetry
$\Spin(5,1)\times \Spin(5,1)$ and there is the possibility of twisting
the whole of the conformal symmetry with the whole of the R-symmetry
group, giving a theory manifestly invariant under the diagonal
$\Spin(5,1)$ subgroup.  One of our aims here is to present this
conformal twisting.  There are some subtleties arising as the
conformal group is non-linearly realised.  Our approach will be to
start with the B-model in which the $\Spin(4)$ Lorentz symmetry is
twisted with a $\Spin(4)$ subgroup of the R-symmetry. This theory is
however not invariant under the full conformal group
\cite{Marcus}---it is invariant under dilatations but not under
special conformal transformations.  We will find modifications of the
usual special conformal transformations that are a symmetry of the
$\SO(4)$ twisted theory, and use these to construct further twistings,
resulting in a topological conformal field theory with a BRST charge
and an anti-BRST charge, both of which are invariant under the twisted
conformal group.
 
This paper is organised as follows.  In Section~\ref{sec:group} we
will discuss the group theory behind the conformal twisting of the
Euclidean $\Neq4$ supersymmetric Yang--Mills theory.  We also discuss
the topological conformal field theories arising from twisted $\Neq2$
supersymmetric theories at conformal fixed points.  In
Section~\ref{sec:conformal} we discuss the $\SO(4)$ twisted Euclidean
$\Neq4$ theory and its symmetries, and in particular find new
modifications of the standard special conformal transformations that
are a symmetry of the theory.  We find linear combinations of
(modified) conformal transformations and R-symmetries that constitute
the twisted $\Spin(5,1)$ symmetry of this theory.  These twisted
conformal generators do not commute with the scalar supercharges, but
rather yield extra conformal supercharges.  We will then show that two
linear combinations of the supercharges are conformally invariant and
define an anticommuting pair of conformally invariant BRST operators.
The conformal symmetry in the twisted theory does not act in the
standard way.  We partially resolve this in Section~\ref{sec:fields}
by redefining the fields in such a way that the conformal
transformations take a more standard form, giving a new form of the
action invariant under the twisted conformal group and BRST charges.

In Section~\ref{sec:currents}, we construct the conserved BRST-exact
symmetric traceless energy-momentum tensor of the flat space twisted
theory.  Then, in Section~\ref{sec:gravity} we couple the theory to
gravity, adding non-minimal terms to the minimally-coupled action to
obtain a theory which is invariant under Weyl rescalings and two BRST
symmetries, with an action that is BRST-exact.  This allows us to
briefly discuss the topological invariants arising as observables in
the theory.

In Section~\ref{sec:thet} we discuss the theta-term in the action and
S-duality.  Finally, in Section~\ref{sec:holography} we discuss some
of the implications of our results to the arguments of
\cite{Hull-Timelike} that such a topological conformal field theory
should have a holographic description as a theory in 5-dimensional de
Sitter space.

We also mention here some other work, not directly related to ours, that has
been done on topological field theories and holography in recent
years.  A version of topological holography in three dimensions has
been developed. On the gauge theory side this involves Chern--Simons
theory, describing knot and three manifold invariants.  There is a
description of this theory using open topological A-strings ending on
a Lagrangian submanifold of a Calabi-Yau threefold \cite{WittenCSST}.
In \cite{GopaVafa} it was proposed for the case of the three sphere
that there is a dual formulation of this Chern--Simons theory based on
closed topological A strings on the resolved conifold.  This proposal
has since been elaborated and extended (see \cite{LabastidaMarino} and
references therein).  A four dimensional topological field theory and
its possible holographic dual has also been discussed in
\cite{Husain}.


\section{Twisting and  group theory}
\label{sec:group}

Dimensional reduction of ($9{+}1$)-dimensional supersymmetric
Yang--Mills from $\RR^{9,1}$ to $\RR^4$ gives the four dimensional
Euclidean $\Neq4$ supersymmetric Yang--Mills theory.  The
ten-dimensional theory has sixteen supercharges in a real chiral
representation $\drep{16}$ of the spin group $\Spin(9,1)$.  After
dimensional reduction, the R-symmetry group is $\Spin(5,1)$, which is
isomorphic to $\SU^*(4) \cong \SL(2,\HH)$, with right and left handed
complex Weyl spinor representations of complex dimension $4$, which we
will denote $\drep{4}$ and $\drep{4'}$ respectively.  Under
$\Spin(9,1) \supset \SU(2) \times \SU(2) \times \SU^*(4)$, the spinors
decompose as
\begin{equation}
  \drep{16} \to (\drep{2},\drep{1},\drep{4'}) \oplus
  (\drep{1},\drep{2},\drep{4}).
\end{equation}
We introduce indices $I,J=1,...,4$ labelling the $\drep{4}$ of
$\SU^*(4)$, $I',J'=1,...,4$ for the $\drep{4'}$ of $\SU^*(4)$,
$A,B=1,2$ labelling the $\drep{2}$ of the first $\SU(2)$ and $\dot
A,\dot B=1,2$ for the $\drep{2}$ of the second $\SU(2)$.  All of these
indices are raised or lowered by complex conjugation,

One must take the underlying real representation of a complex
representation $R$ with a real structure: the Majorana condition on a
spinor in the $\drep{16}$ of $\Spin (9,1)$ becomes a symplectic
Majorana condition in four dimensions. For example, the spinor $
\lambda _{AI'}$ in the $(\drep{2},\drep{1},\drep{4'})$ arising from
the reduction of a Majorana-Weyl fermion in 10 dimensions satisfies
the reality condition
\begin{equation}
  \label{eq:real}
(\lambda ^*) ^{AI'} = \epsilon ^{AB} \Omega ^{I'J'} \lambda _{BJ'},
\end{equation}
where $\Omega ^{I'J'} $ is the symplectic invariant of $\SU^*(4)$
(which can be thought of as the charge conjugation matrix in 5+1
dimensions).

On dimensional reduction, the ten-dimensional super-Poincaré
generators $\sM_{MN}$, $\sP_M$ ($M,N = 1,...10$) and $\sQ$ decompose
into the generators $\sM_{mn}$ and $\sP_m$ ($m,n =1,...,4$) of the
four-dimensional Euclidean group $\ISO(4)$, the $\SO(5,1)$ R-symmetry
generators $\sR_{\mu\nu}$ (antisymmetric in $\mu,\nu$, with
$\mu,\nu=1,...,6$), and the supercharges $\sQ_{I'A}$ and $\sQ_{I \dot
  A}$.  The supercharge $\sQ_{I'A}$ transforms in the
$(\drep{2},\drep{1},\drep{4'})$ representation of $\SU(2) \times
\SU(2) \times \SU^*(4)$, whilst $\sQ_{I\dot A}$ transforms in the
$(\drep{1},\drep{2},\drep{4})$.

The four-dimensional theory is superconformally invariant, with
superconformal group $\SU^*(4|4)$ (a different real form of
$\SU(2,2|4)$) generated by the super-Poincaré generators together with
the dilatation $\sD$, the special conformal generator $\sK_m$ and the
conformal supercharges $\sS_{I'\dot A}$ and $\sS_{IA}$.  The bosonic
subgroup is $\Spin(5,1) \times \Spin(5,1) \cong \SU^*(4) \times
\SU^*(4)$, the product of the Euclidean conformal group and the
R-symmetry.  The 32 (conformal) supercharges transform in the
$(\drep{4'},\drep{4}) \oplus (\drep{4},\drep{4'})$ representation of
$\SU^*(4) \times \SU^*(4)$, with a symplectic Majorana condition using
the symplectic invariants of both factors.

The \emph{conformal} twisting is the diagonal embedding $g\mapsto
(g,g)$ of the conformal group $\SU^*(4)$ in the bosonic symmetry
$\SU^*(4) \times \SU^*(4)$. Under the diagonal embedding, we have
\begin{equation}
  \label{eq:44'}
    [(\drep{4'},\drep{4})] \to \drep{15} \oplus \drep{1}
    \qquad\text{and}\qquad
    [(\drep{4},\drep{4'})] \to \drep{15} \oplus \drep{1}~,
\end{equation}
yielding two scalar supercharges. The bosonic generators of
$\SU^*(4)\times \SU^*(4)$ give under the embedding $
(\drep{15},\drep{1}) \to \drep{15}$ and $ (\drep{1},\drep{15}) \to
\drep{15}$.  The bracket of any two fermionic scalar supercharges must
be a scalar bosonic generator.  As there are none, we conclude that
the scalar supercharges anticommute with each other and each squares
to zero.  In other words, the twisted theory has two (anticommuting)
BRST operators.  After the twisting, one has symmetry under the
diagonal subgroup $\SU^*(4)_D$, which we will refer to as the twisted
conformal group.

The conformal group $\SU^*(4)$ is non-linearly realised, but there is
a subgroup $\CSpin(4):= \SU(2) \times \SU(2) \times \RR^+ \subset
\SU^*(4)$, generated by the spin group and the dilatations, which is a
manifest symmetry of the Euclidean $\Neq4$ supersymmetric Yang--Mills
theory.  In particular, all fields transform as irreducible
representations of $\CSpin(4) \times \SU^*(4)$.

The generators of the original superconformal algebra are
\begin{equation}
  \{\sP_{m}, \sM_{mn}, \sD, \sK_{m}, \sR_{\mu \nu}, \sQ_{I' A}, \sQ_{I\dot
A},  \sS_{IA}, \sS_{I'\dot A} \},
\end{equation}
transforming in the following representation of $G = \SU(2) \times
\SU(2) \times \RR^+ \times \SU^*(4)$:
\begin{multline*}
  (\drep{2},\drep{2},\drep{1})^{+2}\oplus
  (\drep{3},\drep{1},\drep{1})^{0} \oplus
  (\drep{1},\drep{3},\drep{1})^{0} \oplus
  (\drep{1},\drep{1},\drep{1})^{0} \oplus
  (\drep{2},\drep{2},\drep{1})^{-2}\\
  \oplus (\drep{1},\drep{1},\drep{15})^{0} \oplus
  (\drep{2},\drep{1},\drep{4})^{-1} \oplus
  (\drep{1},\drep{2},\drep{4'})^{-1} \oplus
  (\drep{2},\drep{1},\drep{4'})^{+1} \oplus
  (\drep{1},\drep{2},\drep{4})^{+1}~,
\end{multline*}
with the superscript indicating the $\RR^+$ conformal grading.  In
turn, the generators $\sR_{\mu \nu}$ of the $\SU^*(4)$ R-symmetry
transform as the $\drep{15}$ of $\SU^*(4)$, which breaks into the
following representations of the subgroup $\SU(2) \times \SU(2) \times
\RR^+$:
\begin{equation}
  (\drep{2},\drep{2})^{+2}\oplus
  (\drep{3},\drep{1})^{0} \oplus
  (\drep{1},\drep{1})^{0} \oplus
  (\drep{1},\drep{3})^{0} \oplus
  (\drep{2},\drep{2})^{-2}.   
\end{equation}
Introducing indices $m',n'=1,...,4$ for the $\SO(4)$ subgroup of
$\SO(5,1)$, the R-symmetry generators then decompose as
\begin{equation}
  \sR_{\mu\nu} \mapsto \{
\ssp_{m'}, \sm_{m'n'}, \sd,
  \ssk_{m'}
\}~,
\end{equation}
with the $\SU(2) \times \SU(2) \times \RR^+$ generated by $\sm_{m'n'}$
and $\sd$.  The $\RR^+$ gradings of the generators $\{ \ssp_{m'},
\sm_{m'n'}, \sd, \ssk_{m'}\}$ are thus $\{2,0,0,-2\}$.

The $\Spin(4)$ twisting to give the B-model of \cite{Marcus} is
achieved by twisting the action of the rotation generators $\sM_{mn}$
with the action of the R-symmetry generators $\ssm_{m'n'}$, so that
the resulting theory is manifestly invariant under the new spin
generators defined by
\begin{equation}
  \label{eq:dsflkd}
  \mathbb{M}_{mn} \equiv \sM_{mn} +
  \sm_{mn},
\end{equation}
with the indices $m,n$ identified with $m',n'$.  This can then be
enhanced to a $\CSpin(4)$ twisting by twisting the action of the
dilatation $\sD$ with the action of the $\ssd$ ghost number $\RR^+$
symmetry, to obtain a twisted dilatation generator
\begin{equation}
  \label{eq:sflj}
  \mathbb{D} \equiv \sD + \sd.
\end{equation}
Then the twisted conformal weight of a field is the sum of the
conformal weight (defined as half the $\RR^+$ conformal grading) and
ghost number.

To obtain the full conformal twisting, one must in addition twist the
momenta $\sP_m$ and the special conformal generators $\sK_m$ by the
appropriate R-symmetry generators, so that the twisted conformal
generators are
\begin{equation}
\label{eq:twistgens}
  \begin{aligned}[t]
    \mathbb{P}_m &\equiv \sP_m +   \ssp_m \\
    \mathbb{M}_{mn} &\equiv \sM_{mn} +
    \sm_{mn}\\
    \mathbb{D} &\equiv \sD + \sd \\
    \mathbb{K}_m &\equiv \sK_m +  
    \ssk_m~.
  \end{aligned}
\end{equation}

Under the $\CSpin(4)$ twist, the spinor supercharges, which transform
as the $(\drep{2},\drep{1},\drep{4'})^{+1}\oplus
(\drep{1},\drep{2},\drep{{4}})^{+1}$ of $\CSpin(4)\times \SU^*(4)$,
are twisted to generators in the following representations of the
diagonal $\CSpin(4)$:
\begin{equation}
  (\drep{2},\drep{1},\drep{4'})^{+1}\oplus
  (\drep{1},\drep{2},\drep{{4}})^{+1} \to (\drep{3},\drep{1})^{0}\oplus
  (\drep{1},\drep{1})^{0}\oplus (\drep{2},\drep{2})^{+2}\oplus
  (\drep{2},\drep{2})^{+2}\oplus (\drep{1},\drep{3})^{0}\oplus
  (\drep{1},\drep{1})^{0}.  
\end{equation}
This corresponds to the replacements $\{ \sQ_{I' A}, \sQ_{I\dot A}\}
\to \{\sQ^{\pm (0)}_{[mn]}, \sQ^{(0)}, \tilde{\sQ}^{(0)},
\sQ^{(+2)}_m, {\tilde{\sQ}}^{(+2)}_m \}$ and similarly one has
$\{\sS_{I'\dot A}, \sS_{IA}\} \to \{\sS^{\pm (0)}_{[mn]}, \sS^{(0)},
\tilde{\sS}^{(0)}, \sS^{(-2)}_m, {\tilde{\sS}}^{(-2)}_m \}$.
Using this then allows the original superconformal algebra to be
decomposed in terms of brackets involving the twisted supercharges.
We are primarily interested in the scalar supercharges here (ie the
singlets $\sQ,\tilde\sQ,\sS,\tilde\sS$) and their algebra is found to
be (dropping the superscripts denoting the $\RR^+$ conformal gradings)
\begin{equation}
  \begin{aligned}[m]
    [\sQ,{\mathbb{P}}_m ] &= -\sQ_m \\
    [\sQ,{\mathbb{K}}_m ] &= -\tilde{\sS}_m\\
    [\sS,{\mathbb{P}}_m ] &= \tilde{\sQ}_m \\
    [\sS,{\mathbb{K}}_m ] &= -\sS_m\\
    [\sQ,\sS] &= -4 \mathbb{D}
  \end{aligned}
\quad\quad\quad
 \begin{aligned}[m]
   [\tilde\sQ,{\mathbb{P}}_m ] &= \tilde\sQ_m \\
    [\tilde\sQ,{\mathbb{K}}_m ] &= -{\sS}_m\\
    [\tilde\sS,{\mathbb{P}}_m ] &= {\sQ}_m \\
    [\tilde\sS,{\mathbb{K}}_m ] &= \tilde\sS_m\\
    [\tilde\sQ,\tilde\sS] &= -4 \mathbb{D}
\end{aligned}
\end{equation}
where we have only written down the non-zero brackets.  The linear
combinations
\begin{equation}
\mathbb{Q}:= \sQ+\tilde{\sS}, \qquad
\tilde{\mathbb{Q}}:= \tilde{\sQ}- \sS
\end{equation}
square to zero, anticommute with each other, and commute with the
twisted conformal generators:
\begin{equation}
  [{\mathbb{Q}}, {\mathbb{X}}] = [\tilde{\mathbb{Q}}, \mathbb{X}] = 0 
\qquad
  \text{for any}\quad \mathbb{X} \in \{\mathbb{P}_m, \mathbb{M}_{mn},
  \mathbb{D},\mathbb{K}_m,\mathbb{Q},\tilde{\mathbb{Q}}\}~.
\end{equation}
The $\Spin(5,1)$ generators $\{\mathbb{P}_m, \mathbb{M}_{mn},
\mathbb{D},\mathbb{K}_m\}$ satisfy the algebra
\begin{equation}
\label{eq:so51algebra}
  \begin{aligned}[t]
    [\MM_{mn}, \MM_{pq}] &=  \delta_{mp} \MM_{nq} + \delta_{nq} \MM_{mp}
             -\delta_{mq} \MM_{np} - \delta_{np} \MM_{mq} \\
    [\MM_{mn}, \PP_p] &=  \delta_{mp} \PP_n - \delta_{pn} \PP_m\\  
    [\MM_{mn}, \KK_p] &=  \delta_{mp} \KK_n - \delta_{pn} \KK_m\\
    [\DD, \PP_m ] &=  \PP_m\\
    [\DD, \KK_m ] &= - \KK_m \\
    [\KK_m, \PP_n ] &=  2\delta_{mn} \DD + 2 \MM_{mn}~,
  \end{aligned}
\end{equation}
with all other brackets vanishing.

It is interesting to ask how much of this structure survives 
for twisted $\Neq2$ theories at conformal
fixed points. Such theories have been investigated in
\cite{Mooreand,Mooreand2}.  A twisted $\Neq2 $ theory is invariant
under Poincar\' e symmetry, together with the symmetries generated by
a BRST charge $\sQ$ and a ghost number charge $\sd$. At a conformal
fixed point, the theory is invariant under the conformal group, which is
generated by the twisted Lorentz generators $\mathbb{M}_{mn}$ together with
$\{ \sP_m,  \sD ,\sK_m\}$, and also invariant under
the ghost number symmetry generated by $\sd$, the  BRST symmetry
generated by $\sQ$ and a conformal BRST generated by $\sS$,
arising from twisting the conformal supersymmetry. These satisfy a
subalgebra of
the algebra discussed above, but without the generators $\tilde \sQ,
\tilde \sS$. 
In this case, there do not seem to be any further twistings that lead to
conformally invariant BRST operators of the type discussed above.


\section{The diagonally twisted $\Neq4$  theory and conformal invariance}
\label{sec:conformal}

Before twisting, the fields of the Euclidean $\Neq4$ supersymmetric
Yang--Mills theory are $A_{m}, {\lambda} _{IA}, {\lambda} _{I' \dot
  A}, {\phi}_{IJ}$, with $\phi$ antisymmetric in its indices.
Consider first the diagonal twisting of the $\SO(4)$ Lorentz symmetry
with an $\SO(4)$ subgroup of the R-symmetry to obtain the B-model.
The vector field is an R-singlet and remains unchanged, while the 6
scalars $ \phi_{IJ}$ twist to a vector $V_m$ and two scalars $B,C$.
The fermions ${\lambda}$ twist to give the anticommuting fields
$\psi_m,\tilde\psi_m,\chi^{\pm}_{mn},\eta,\tilde\eta$, where
$\chi^{\pm}_{mn}$ are 2-forms satisfying $\chi^{\pm}_{mn}=\pm
*\chi^{\pm}_{mn}$. We define $(X_{[mn]})^{\pm}\equiv \half(X_{[mn]}
\pm *X_{[mn]})$ and $X_{[mn]}\equiv \half(X_{mn}-X_{nm})$.

The $\Spin(4)$ twisted action is \cite{Marcus}:
\begin{equation}
\label{eq:S0Lozano}
  \begin{split}
    \mathcal{S}^{(0)} &= \frac{1}{e^2}\int d^4 x\,
    \tr \biggl( -\eD_m B\eD^m C -\eD_m V_n\eD^m V^n
    -\tfrac14 F_{mn} F^{mn} \\
    & {} +\eD_m\psi_n (4\chi^{+ mn} - \delta^{mn} \eta) +
    \eD_m\tilde\psi_n (4\chi^{- mn} - \delta^{mn} \tilde\eta) \\
    & {} -\tfrac{i}{8\sqrt{2}}\,((4\chi^{+}_{mn} -\delta_{mn}
    \eta)[4\chi^{+mn}-\delta^{mn} \eta,C] +(4\chi^-_{mn}
    -\delta_{mn}\tilde\eta)[4\chi^{-mn} -\delta^{mn} \tilde\eta,C])\\
    &{} -i\sqrt{2}\,((4\chi^{+}_{mn} -\delta_{mn} \eta)[\tilde\psi^{m} ,
    V^{n}] - (4\chi^-_{mn} -\delta_{mn} \tilde\eta)[\psi^{m} ,
    V^{n}]) \\
    &{} +i\sqrt{2}\,(\psi_m[\psi^m,B] + \tilde\psi_m[\tilde\psi^m,B]) \\
    &{} -\half[B,C]^2 +2[B,V_m][C,V^m]+ [V_m,V_n] [V^m,V^n]
    \biggr) \\
    &{} -\frac{i\theta}{ 32\pi^2}\int d^4 x\,\tr *F_{mn} F^{mn}~.
  \end{split}
\end{equation}
We follow the notation in   \cite{Lozano}.  Also
$\eD_{m} := \d_{m} + i[A_{m},\,]$ is the gauge covariant derivative
acting on the fields, which are valued in the adjoint representation
of the Lie algebra of the non-abelian gauge symmetry group. The field
strength is
\begin{equation}
  F_{mn} = \d_{m} A_{n} - \d_{n} A_{m} +i [A_{m},A_{n}],
\end{equation}
while $e$ is the usual Yang--Mills coupling constant and
$\theta$ is the theta-parameter, which will be discussed 
further in section 7.

The theory is invariant under the $\SO(1,1)$ scale transformations
generated by the dilatation $\sD$, with each field having a conformal
weight $c$, and under the $\SO(1,1)$ subgroup of the $\SO(5,1)$
R-symmetry group that commutes with the $\SO(4)$ that was twisted.
This is generated by $\ssd$ and is usually referred to as the
ghost-number symmetry, with the $\SO(1,1)$ weight of each field
referred to as the ghost number $g$ of that field.  The fields
\begin{equation}
 \{B,C,A_m,V_m;\psi_m,\tilde\psi_m,\chi^{\pm}_{mn},\eta,\tilde\eta\}
\end{equation}
have conformal weights and ghost numbers $\{(c,g)\}$ given by
\begin{equation}
  \{(1,1), (1,-1), (1,0), (1,0) ; (3/2,-1/2), (3/2,-1/2), (3/2,1/2),
  (3/2,1/2), (3/2,1/2)\},
\end{equation}
respectively.  The twisted dilatation was defined as $\mathbb{D}
\equiv \sD + \sd$ and for a field with $\sD, \sd$ weights $\{(c,g)\}$,
the weight with respect to $\mathbb{D} $ is $c+g$.  Then the twisted
bosonic fields are
\begin{equation}
  \{B^{(2)},C^{(0)},A^{(1)}_m,V^{(1)}_m\},  
\end{equation}
and the fermionic fields are
\begin{equation}
  \{\psi^{(1)}_m , \tilde\psi^{(1)}_m, \chi^{(2)\pm}_{mn},
  \eta^{(2)}, \tilde\eta^{(2)}\},
\end{equation}
where the superscripts (suppressed in the following) give the twisted
conformal weight $c+g$.  The twisted theory is automatically invariant
under the twisted scale transformations generated by $\mathbb{D} $ and
so can be regarded as a twisting of $\CSpin(4)$ rather than just of
$\Spin(4)$.  The twisted dilatations act in the standard way, i.e.,
$\mathbb{D} \cdot \Phi = (x^n\partial_n + (c+g)_\Phi)\Phi$, with the
weight $(c+g)_\Phi$ for each field $\Phi$ defined as above.

As discussed in the previous section, twisting produces various
fermionic charges. In particular, we will be interested in the scalar
BRST supercharges $\sQ$ and $\tilde{\sQ}$. The action of these on the
fields is given by
\begin{equation}
     \label{eq:BRST}
  \begin{aligned}[m]
    &\sQ\cdot A_m =  2\psi_m \\
    &\sQ\cdot \psi_m =  \sqrt{2}\,\eD_m C  \\
    &\sQ\cdot \tilde\psi_m = -2i[V_m,C] \\
    &\sQ\cdot \chi^{+}_{mn} = - F^{+}_{mn} +2i [V_m,V_n]^{+}  \\
    &\sQ\cdot \chi^-_{mn} = 2\sqrt{2}\,(\eD_{[m}V_{n]})^- \\
    &\sQ\cdot \eta  = 2i[B,C] \\
    &\sQ\cdot \tilde\eta = -2\sqrt{2}\,\eD_m V^m \\
    &\sQ\cdot B = \sqrt{2}\,\eta  \\
    &\sQ\cdot C = 0 \\
    &\sQ\cdot V_m = -\sqrt{2}\,\tilde\psi_m
  \end{aligned}
  \qquad
  \begin{aligned}[m]
    &\tilde{\sQ}\cdot A_m =  -2\tilde\psi_m \\
    &\tilde{\sQ}\cdot \psi_m =  -2i[V_m,C] \\
    &\tilde{\sQ}\cdot \tilde\psi_m = -\sqrt{2}\,\eD_m C\\
    &\tilde{\sQ}\cdot \chi^{+}_{mn} = 
    2\sqrt{2}\,(\eD_{[m}V_{n]})^{+}\\
    &\tilde{\sQ}\cdot \chi^-_{mn} =  F^-_{mn} -2i[V_m,V_n]^-\\
    &\tilde{\sQ}\cdot \eta = -2\sqrt{2}\,\eD_m V^m\\
    &\tilde{\sQ}\cdot \tilde\eta = -2i[B,C]\\
    &\tilde{\sQ}\cdot B = -\sqrt{2}\,\tilde\eta\\
    &\tilde{\sQ}\cdot C = 0\\
    &\tilde{\sQ}\cdot V_m = -\sqrt{2}\,\psi_m.
  \end{aligned}
\end{equation}
The associated infinitesimal transformations are given as usual by
$\delta_Q X \equiv \epsilon \sQ \cdot X$ and $\delta_{\tilde{\sQ}} X
\equiv \tilde\epsilon \tilde{\sQ} \cdot X$, where
 $\epsilon$ and $\tilde\epsilon$ are the corresponding
anti-commuting scalar parameters. The BRST operators $\sQ$ and
$\tilde{\sQ}$ are nilpotent and anticommute with each other, up to
gauge transformations and on-shell (utilising the $\chi^\pm, \eta$ and
$\tilde\eta$ equations of motion from \eqref{eq:S0Lozano}). The
on-shell condition may be removed by introducing auxiliary fields in a
standard way \cite{Lozano}.

The other symmetries of the theory include the following.  First we
have the twisted rotations, which act in the usual way:
\begin{equation}
  \begin{aligned}[t]
    &\MM_{mn} \cdot A_p =  (x_m\d_n-x_n\d_m) A_p +
    \delta_{mp} A_n - \delta_{np} A_m \\
    &\MM_{mn} \cdot \psi_p =  (x_m\d_n-x_n\d_m) \psi_p
    + \delta_{mp} \psi_n  - \delta_{np} \psi_m \\
    &\MM_{mn} \cdot \tilde\psi_p = 
    (x_m\d_n-x_n\d_m)\tilde\psi_p + \delta_{mp} \psi_n -
    \delta_{np} \tilde\psi_m \\
    &\MM_{mn} \cdot \chi^{\pm}_{pq} = 
    (x_m\d_n-x_n\d_m)\chi^{\pm}_{pq} - \delta_{mp} \chi^{\pm}_{nq}  
    - \delta_{np} \chi^{\pm}_{mq} + \delta_{mq} \chi^{\pm}_{pn} -
      \delta_{nq} \chi^{\pm}_{pm} \\  
    &\MM_{mn} \cdot \eta =  (x_m\d_n-x_n\d_m)\eta \\  
    &\MM_{mn} \cdot \tilde\eta =  (x_m\d_n-x_n\d_m)\tilde\eta  \\
    &\MM_{mn} \cdot B =  (x_m\d_n-x_n\d_m) B  \\
    &\MM_{mn} \cdot C =  (x_m\d_n-x_n\d_m) C  \\
    &\MM_{mn} \cdot V_p =  (x_m\d_n-x_n\d_m) V_p +
    \delta_{mp} V_n - \delta_{np} V_m~.
  \end{aligned}
\end{equation}
The spacetime translations act in the standard way,
\begin{equation}
 \sP_m \cdot\Phi=   \d_m \Phi
\end{equation}
for all fields $\Phi$.

The R-symmetry generators $\ssp_m, \ssk_m$ lead to the following
symmetries of the twisted theory:
\begin{equation}
  \begin{aligned}
    \ssp_m \cdot \psi_p& =   - \half  
    (4\chi^-_{mp}+\delta_{mp} \tilde\eta) \\
    \ssp_m \cdot \tilde\psi_p& =    \half 
    (4\chi^{+}_{mp}+\delta_{mp} \eta) \\
       \ssp_m \cdot C& =   2 V_m \\
    \ssp_m \cdot V_p& =    -  \delta_{mp} B~,
  \end{aligned}
\end{equation}
and, using    $\delta_{\ssk}$
defined  by $\delta_{\ssk} X =  \kappa^m \ssk_m \cdot X$,
\begin{equation}
  \begin{aligned}[t]
      &\delta_{\ssk}\chi^+_{pq} = - 2
      (\kappa_{[p}\tilde\psi_{q]})^+ \\
    &\delta_{\ssk}\chi^-_{pq} =   2(\kappa_{[p}psi_{q]})^- \\
   &\delta_{\ssk} \eta =  - 2  \kappa^m \tilde\psi_m \\
    &\delta_{\ssk} \tilde\eta =   2  \kappa^m \psi_m \\
    &\delta_{\ssk} B =   2  \kappa^m V_m \\
     &\delta_{\ssk} V_p = -  \kappa_p C~.
  \end{aligned}  
\end{equation}

The original $\Neq4$ theory was conformally invariant, but the twisted
theory is invariant under scale transformations generated by $\sD$,
but not under the conformal boosts in which $\sK_m$ acts in the
standard way \cite{Marcus}, so that in this sense it is scale
invariant but not conformally invariant.  The trace of the stress
tensor is non-zero, but is a total derivative, so that the integral
over $\mathbb{R} ^4$ of the trace of the stress tensor vanishes (with
suitable boundary conditions), signalling dilatation invariance.
However, we have found some modifications of the action of the action
of $\sK_m$ that are a symmetry of the action \eqref{eq:S0Lozano}.
These are conveniently written using $\delta_{\sK}$, defined by
$\delta_{\sK} X = \kappa^m \sK_m \cdot X$.  One finds
\begin{equation}
  \begin{aligned}[t]
    &\delta_{\sK} A_p =  \kappa^m (2x_m x \cdot \d - x^2 \d_m +2 x_m)
    A_p + 4 \kappa_{[p}x_{q]}A^q \\
    &\delta_{\sK} \psi_p =  \kappa^m (2x_m x \cdot \d - x^2 \d_m + 3
    x_m) \psi_p +
    4(\kappa_{[p}x_{q]})^- \psi^q     \\
    &\delta_{\sK} \tilde\psi_p =  \kappa^m (2x_m x \cdot \d - x^2 \d_m  
    + 3 x_m) \tilde\psi_p + 4 (\kappa_{[p}x_{q]})^+ \tilde\psi^q \\
    &\delta_{\sK}\chi^+_{pq} =  \kappa^m (2x_m x \cdot \d - x^2 \d_m
    + 2 x_m) \chi^+_{pq} -4 (\kappa_{[p}x^k\chi^+_{q]k})^+ -
      (\kappa_{[p} x_{q]})^+ \eta  \\
    &\delta_{\sK}\chi^-_{pq} =  \kappa^m (2x_m x \cdot \d - x^2 \d_m
    +2 x_m) \chi^-_{pq} -4 (\kappa_{[p}x^k\chi^-_{q]k})^- -
      (\kappa_{[p} x_{q]})^-
      \tilde\eta    \\
    &\delta_{\sK} \eta =  \kappa^m (2x_m x \cdot \d - x^2 \d_m +3 x_m)
    \eta
    + 4 \kappa^m x^n\chi_{mn}^+  \\
    &\delta_{\sK} \tilde\eta =  \kappa^m (2x_m x \cdot \d - x^2 \d_m
    +3 x_m) \tilde\eta
    + 4 \kappa^m x^n\chi_{mn}^-   \\
    &\delta_{\sK} B =  \kappa^m (2x_m x \cdot \d - x^2 \d_m +2 x_m)B
    \\
    &\delta_{\sK} C =  \kappa^m (2x_m x \cdot \d - x^2 \d_m +2 x_m) C \\
    &\delta_{\sK} V_p =  \kappa^m (2x_m x \cdot \d - x^2 \d_m +2 x_m)
    V_p .
  \end{aligned}  
\end{equation}
These are not standard conformal boost transformations due to the
presence of extra terms, some of which mix different fields.

The algebra of these symmetries does not close to give the conformal
algebra \eqref{eq:so51algebra}, and in particular the commutator of
$\sP$ and $\sK$ is not of the right form.  However, from the
discussion of the previous section, we are interested in the full
$\Spin(5,1)$ twisting in which the full conformal group is twisted
with the full R-symmetry group, and in particular this means twisting
$\sP$ with $\ssp$ and $\sK$ with $\ssk$.  We therefore consider the
symmetries generated by
\begin{equation}
\label{twistgens2}
  \begin{aligned}[t]
    \mathbb{P}_m &\equiv \sP_m +  \mu  \ssp_m \\
    \mathbb{K}_m &\equiv \sK_m + \mu ^{-1} 
    \ssk_m~,
  \end{aligned}
\end{equation}
where we have included an arbitrary parameter $\mu$.  Notice that the
scalings $\ssp\rightarrow \mu\ssp, \ssk\rightarrow \frac{1}{\mu}\ssk$
leave invariant the $\SO(5,1)$ R-symmetry algebra generated by $\{
\ssp, \ssk, \ssm, \ssd \} $.

For these new translations we thus have
\begin{equation}
  \begin{aligned}
    \PP_m \cdot A_p &=   \d_m A_p \\
    \PP_m \cdot \psi_p& =  \d_m \psi_p - \half \mu
    (4\chi^-_{mp}+\delta_{mp} \tilde\eta) \\
    \PP_m \cdot \tilde\psi_p& =  \d_m \tilde\psi_p + \half \mu
    (4\chi^{+}_{mp}+\delta_{mp} \eta) \\
    \PP_m \cdot \chi^{\pm}_{pq}& =   \d_m \chi^{\pm}_{pq} \\
    \PP_m \cdot \eta& =   \d_m \eta \\
    \PP_m \cdot \tilde\eta& =  \d_m \tilde\eta \\
    \PP_m \cdot B& =  \d_m B \\
    \PP_m \cdot C& =   \d_m C + 2\mu V_m \\
    \PP_m \cdot V_p& =  \d_m V_p - \mu \delta_{mp} B~,
  \end{aligned}
\end{equation}
and for the new special conformal transformations, using $\delta_\KK$
defined by $\delta_\KK X = \kappa^m \KK_m \cdot X$, we have
\begin{equation}
  \begin{aligned}[t]
    &\delta_\KK A_p =  \kappa^m (2x_m x \cdot \d - x^2 \d_m +2 x_m)
    A_p + 4 \kappa_{[p}x_{q]}A^q \\
    &\delta_\KK \psi_p =  \kappa^m (2x_m x \cdot \d - x^2 \d_m + 3
    x_m) \psi_p +
    4(\kappa_{[p}x_{q]})^- \psi^q     \\
    &\delta_\KK \tilde\psi_p =  \kappa^m (2x_m x \cdot \d - x^2 \d_m  
    + 3 x_m) \tilde\psi_p + 4 (\kappa_{[p}x_{q]})^+ \tilde\psi^q \\
    &\delta_\KK\chi^+_{pq} =  \kappa^m (2x_m x \cdot \d - x^2 \d_m
    + 2 x_m) \chi^+_{pq} -4 (\kappa_{[p}x^k\chi^+_{q]k})^+ -
      (\kappa_{[p} x_{q]})^+ \eta - 2\mu^{-1}
      (\kappa_{[p}\tilde\psi_{q]})^+ \\
    &\delta_\KK\chi^-_{pq} =  \kappa^m (2x_m x \cdot \d - x^2 \d_m
    +2 x_m) \chi^-_{pq} -4 (\kappa_{[p}x^k\chi^-_{q]k})^- {-
      (\kappa_{[p} x_{q]})^-
      \tilde\eta + 2 \mu^{-1} (\kappa_{[p} \psi_{q]})^-}    \\
    &\delta_\KK \eta =  \kappa^m (2x_m x \cdot \d - x^2 \d_m +3 x_m)
    \eta
    + 4 \kappa^m x^n\chi_{mn}^+ - 2 \mu^{-1} \kappa^m \tilde\psi_m \\
    &\delta_\KK \tilde\eta =  \kappa^m (2x_m x \cdot \d - x^2 \d_m
    +3 x_m) \tilde\eta
    + 4 \kappa^m x^n\chi_{mn}^- + 2 \mu^{-1} \kappa^m \psi_m \\
    &\delta_\KK B =  \kappa^m (2x_m x \cdot \d - x^2 \d_m +2 x_m)B
    + 2 \mu^{-1} \kappa^m V_m \\
    &\delta_\KK C =  \kappa^m (2x_m x \cdot \d - x^2 \d_m +2 x_m) C \\
    &\delta_\KK V_p =  \kappa^m (2x_m x \cdot \d - x^2 \d_m +2 x_m)
    V_p - \mu^{-1} \kappa_p C~.
  \end{aligned}  
\end{equation}
One can check that the generators $\{\mathbb{P}_m, \mathbb{M}_{mn},
\mathbb{D},\mathbb{K}_m\}$ as defined above are symmetries of the
action \eqref{eq:S0Lozano} and satisfy the $\Spin(5,1)$ algebra
\eqref{eq:so51algebra} for any $\mu$.


Just as twisting the ordinary supersymmetries of the $\Neq4$ theory
gives a number of fermionic generators, including the two scalar BRST
charges $\sQ,\tilde \sQ$, twisting the conformal supersymmetries also
gives a set of fermionic symmetries including those generated by two
further scalar charges $\sS$ and $\tilde\sS$.  We will call these
\emph{SBRST} generators.  They generate the following symmetries of
the action \eqref{eq:S0Lozano}
\begin{equation}
  \begin{aligned}[t]
    &\sS \cdot A_m =  -x^n  (4 \chi^+_{mn} - \delta_{mn} \eta) \\  
    &\sS \cdot \psi_m =  2x^n  F^-_{mn} - ix_m [B,C] + 4i x^n
    [V_m,V_n]^+ \\
    &\sS \cdot \tilde\psi_m =  \sqrt{2}  [4 x^n (\eD_{[m} V_{n]})^- - 2
    x_{[m} \eD^n V_{n]} +(x^n \eD_n + 2)V_m ] \\
    &\sS \cdot \chi^+_{mn} =  - 2 \sqrt{2}  (x_{[m} \eD_{n]})^+ B \\
    &\sS \cdot \chi^-_{mn} =  4i  [(x_{[m} V_{n]})^-,B] \\
    &\sS \cdot \eta  =  - 2 \sqrt{2}  (x^n \eD_n + 2)B \\
    &\sS \cdot \tilde\eta= -4i  [x^n V_n,B] \\
    &\sS \cdot B = 0 \\
    &\sS \cdot C = -2\sqrt{2}  x^n \psi_n \\
    &\sS \cdot V_m =  \tfrac1{\sqrt{2}} x^n (4 \chi^-_{mn} +
    \delta_{mn} \tilde\eta)
  \end{aligned}
\end{equation}
and
\begin{equation}
  \begin{aligned}[t]
    &\tilde{\sS} \cdot A_m =  x^n  (4 \chi^-_{mn} - \delta_{mn}
    \tilde\eta) \\
    &\tilde{\sS} \cdot \psi_m =  \sqrt{2}  [4 x^n (\eD_{[m} V_{n]})^+ -
    2 x_{[m} \eD^n V_{n]} +(x^n \eD_n + 2)V_m ] \\
    &\tilde{\sS} \cdot \tilde\psi_m =  -2x^n  F^+_{mn} + i x_m [B,C] -
    4i x^n  [V_m,V_n]^- \\
    &\tilde{\sS} \cdot \chi^+_{mn}= 4i  [(x_{[m} V_{n]})^+,B] \\
    &\tilde{\sS} \cdot \chi^-_{mn} = 2\sqrt{2}  (x_{[m}\eD_{n]})^- B \\ 
    &\tilde{\sS} \cdot \eta= -4i  [x^n V_n,B] \\
    &\tilde{\sS} \cdot \tilde\eta  = 2\sqrt{2}  (x^n \eD_n + 2)B \\
    &\tilde{\sS} \cdot B  =  0 \\
    &\tilde{\sS} \cdot C  =  2 \sqrt{2}  x^n \tilde\psi_n \\
    &\tilde{\sS} \cdot V_m =  \tfrac1{\sqrt{2}} x^n  (4 \chi^+_{mn} +
    \delta_{mn} \eta)~.
  \end{aligned}   
\end{equation}
The corresponding infinitesimal transformations are given as usual by
$\delta_S X = \xi \sS \cdot X$ and $\delta_{\tilde S} X = \tilde\xi
\tilde{\sS} \cdot X$, with $\xi$ and $\tilde\xi$ the fermionic scalar
parameters.

The brackets between all the scalar supercharges
$\sQ,\tilde\sQ,\sS,\tilde\sS$ vanish (on-shell and up to gauge
transformations) except for
\begin{equation}
  \label{eq:QSbracket}
  [\sQ,\sS] =  -4\DD~, \qquad  [\tilde \sQ, \tilde\sS] =  -4\DD~.
\end{equation}

The $\SO(4)$ twisted theory with action \eqref{eq:S0Lozano} also has a
discrete $\ZZ_2$ symmetry which acts on both the fields and the
coupling constants in the following way.  It leaves $A_m,B,C,e$
invariant.  It reverses orientation: $\epsilon_{mnpq} \mapsto
-\epsilon_{mnpq}$, so that $\theta \mapsto - \theta$.  Defining
\begin{equation}
  \label{eq:tau}
  \tau :=  \frac{4\pi}{e^2} - i \frac{\theta}{2\pi}~,
\end{equation}
we have that $\tau \mapsto \bar\tau$.  On the remaining fields
and constants, it acts as follows:
\begin{equation}
  \begin{aligned}[t]
    \chi^{\pm}_{mn} & \mapsto -\chi^{\mp}_{mn}\\
    (\psi_m ,\tilde\psi_n ) &\mapsto (-\tilde\psi_m , -\psi_n )\\
    (\eta , \tilde\eta) &\mapsto (-\tilde\eta , -\eta)\\
    V_m &\mapsto -V_m\\
    \mu &\mapsto -\mu~.
  \end{aligned}  
\end{equation}
This symmetry leaves the action \eqref{eq:S0Lozano}
invariant and it exchanges the scalar (S)BRST generators
$(\sQ,\sS) \mapsto (\tilde{\sQ}, \tilde{\sS})$.

Although the theory \eqref{eq:S0Lozano} is conformally invariant and
has four (S)BRST symmetries, it is not the case that the (S)BRST
supercharges are themselves conformally invariant, i.e., they are not
singlets under the action of the $\SO(5,1)$ generated by
$\MM,\PP,\DD,\KK$.  This is clear from \eqref{eq:QSbracket} and the
fact that $\DD$ is not central in the conformal algebra.  Nevertheless
the two linear combinations
\begin{equation}
  \mathbb{Q} :=  \sQ + \mu\tilde{\sS}, \qquad \tilde{\mathbb{Q}} :=
  \tilde{\sQ} -\mu\sS,
\end{equation}
satisfy
\begin{equation}
  \mathbb{Q}^2=0, \qquad \tilde{\mathbb{Q}} ^2=0, \qquad \{
  \mathbb{Q}, \tilde{\mathbb{Q}} \} =0
\end{equation}
(on-shell and up to gauge transformations). We will call these scalar
supercharges \emph{CBRST} charges (the \emph{C} denoting `conformal').
These charges commute with the conformal generators and are mapped
into each other by the $\ZZ_2$ symmetry above.  They are the two
scalar supercharges of the group theoretical analysis in
Section~\ref{sec:group}; see \eqref{eq:44'}.


\section{Redefining the fields}
\label{sec:fields}

In the previous section we have shown that the $\SO(4)$ twisted theory
with action \eqref{eq:S0Lozano} is conformally invariant and moreover
that there are two CBRST symmetries commuting with the conformal
generators.  Nevertheless, the action of the conformal generators does
not take the standard form---for example, the fields do not transform
conventionally under the translations $\PP _m$. It is clear that this
must occur in general in situations like this where one twists the
ordinary translations with a corresponding internal symmetry. This
unconventional behaviour under the translations $\PP _m$ will cause
complications, for example when considering defining the theory on a
curved manifold. It is thus convenient, after having done the twisting
as in the previous section, to {\it undo} the $R$-symmetry
translations $\ssp_m$ by redefining all fields $\Phi$ according to
\begin{equation}
\label{eq:Phione}
\Phi \rightarrow (e^{-\mu x^m\ssp_m})\cdot\Phi.
\end{equation}
Explicitly, this gives the following field redefinitions
\begin{equation}
\label{eq:redefs}
  \begin{aligned}[t]
    V_m & \mapsto V_m + \mu x_m B \\
    C &\mapsto C -2\mu x^m V_m - \mu^2 x^2 B \\
    \psi_m & \mapsto \psi_m - \half \mu x^n (4\chi^-_{mn} -
    \delta_{mn}\tilde\eta) \\
    \tilde\psi_m &\mapsto \tilde\psi_m + \half \mu x^n (4\chi^+_{mn} -  
    \delta_{mn}\eta)~.
  \end{aligned}               
\end{equation}
Note that these redefinitions are consistent with the   $\ZZ_2$
symmetry.

It is straightforward to calculate the effect of the
$\{\mathbb{M}_{mn},\mathbb{P}_m,\mathbb{D}\}$ twisted generators on
the redefined fields.  These are taken as transformations that act on
the fields but not on the space-time coordinates $x^m$.  In
particular, $\PP_m$ does not transform the explicit $x^m$ in the
redefinitions \eqref{eq:redefs}, with the result that its action on
the redefined fields is, by construction, now the standard action
\begin{equation}
\label{eq:newPs}
   \mathbb{P}  _m \cdot \Phi =  \d_m \Phi 
\end{equation}
on any field $\Phi$. Since $\mathbb{M}_{mn}$ and $\mathbb{D}$ commute
with $x^m\ssp_m$, the action of the twisted dilatations and rotations
remains standard
\begin{equation}
 \label{eq:newMDs}
 \begin{aligned}[t]
  \mathbb{M}  _{mn} \cdot \Phi &=  (x_m \d_n - x_n \d_m +\Sigma_{mn}) \Phi \\
    \mathbb{D}     \cdot \Phi &=  (x^n \d_n + \Delta_\Phi)\Phi~,
  \end{aligned}
\end{equation}
where $\Sigma_{mn}$ are the generators of the spin group in the
representation determined by the spin of $\Phi$ and $\Delta_\Phi$ is the
twisted conformal weight of $\Phi$.
However, the special conformal transformations still contain
non-standard $\mu$-dependent terms:
\begin{equation}
\label{eq:newKs}
  \begin{aligned}[t]
    &\delta_{\mathbb{K}} A_p =  
\kappa^m (2x_m x\cdot\d -x^2 \d_m + 2x_m)A_p + 4
    \kappa_{[p}x_{q]}A^q \\
    &\delta_{\mathbb{K}} \psi_p =  
\kappa^m (2x_m x\cdot\d -x^2 \d_m + 2x_m)\psi_p 
    + 4\kappa_{[p}x_{q]}\psi^q \\
    &\delta_{\mathbb{K}} \tilde\psi_p = 
 \kappa^m ( 2 x_m x\cdot\d  - x^2 \d_m + 2
    x_m) \tilde\psi_p + 4 \kappa_{[p}x_{q]}\tilde\psi^q \\
    &\delta_{\mathbb{K}} \chi^+_{pq} = 
 \kappa^m (2x_m x\cdot\d -x^2 \d_m + 4 x_m)
    \chi^+_{pq} - 8 (\kappa_{[p}x^k \chi^+_{q]k})^+ - 2\mu^{-1}
    (\kappa_{[p} \tilde\psi_{q]})^+ \\
    &\delta_{\mathbb{K}} \chi^-_{pq} = 
 \kappa^m (2x_m x\cdot\d -x^2 \d_m + 4 x_m)
    \chi^-_{pq} -8 (\kappa_{[p}x^k \chi^-_{q]k})^- + 2\mu^{-1}
    (\kappa_{[p} \psi_{q]})^- \\
    &\delta_{\mathbb{K}} \eta = 
 \kappa^m (2x_m x\cdot\d -x^2 \d_m + 4x_m) \eta
    - 2\mu^{-1} \kappa^m \tilde\psi_m \\
    &\delta_{\mathbb{K}} \tilde\eta = 
 \kappa^m (2x_m x\cdot\d -x^2 \d_m +
    4x_m)\tilde\eta + 2 \mu^{-1} \kappa^m \psi_m \\
    &\delta_{\mathbb{K}} B = 
 \kappa^m ( 2x_m x\cdot\d -x^2 \d_m + 4x_m) B
    + 2 \mu^{-1} \kappa^m V_m \\
    &\delta_{\mathbb{K}} C = 
 \kappa^m (2 x_m x\cdot\d -x^2 \d_m )C \\
    &\delta_{\mathbb{K}} V_p = 
 \kappa^m (2 x_m x\cdot\d -x^2 \d_m + 2x_m) V_p
    + 4 \kappa_{[p}x_{q]}V^q -\mu^{-1} \kappa_p C~.
  \end{aligned}
\end{equation}
These transformations differ from the usual transformations
\begin{equation}
  \delta_{\mathbb{K}} \Phi =  \kappa^m (2x_m x\cdot\d -x^2 \d_m + 2x_m
  \Delta_\Phi + 2x^n\Sigma_{mn})\Phi~
\end{equation}
by the addition of extra terms proportional to $\mu ^{-1}$.  These
$\mu^{-1}$ terms imply the loss of special conformal symmetry in the
$\mu \to 0$ limit.

In terms of the redefined fields the CBRST symmetry remains simple:
\begin{equation}
 \label{eq:newCBRSTs}
 \begin{aligned}[t]
    &\mathbb{Q} \cdot A_m =  2\psi_m \\
    &\mathbb{Q} \cdot \psi_m =  \sqrt{2}\, \eD_m C \\
    &\mathbb{Q} \cdot \tilde\psi_m = -2i[V_m,C] \\
    &\mathbb{Q} \cdot \chi^+_{mn} =  -F^+_{mn} + 2i[V_m,V_n]^+ \\
    &\mathbb{Q} \cdot \chi^-_{mn} =  2\sqrt{2}\,(\eD_{[m} V_{n]})^- \\
    &\mathbb{Q} \cdot \eta =  2i[B,C] \\  
    &\mathbb{Q} \cdot\tilde\eta =  -2\sqrt{2}\,(\eD_m V^m +2\mu B) \\
    &\mathbb{Q} \cdot B =  \sqrt{2}\, \eta \\
    &\mathbb{Q} \cdot C =  0 \\
    &\mathbb{Q} \cdot V_m =  -\sqrt{2}\, \tilde\psi_m  
  \end{aligned}
  \qquad
  \begin{aligned}[t]
    &\tilde{\mathbb{Q}} \cdot A_m =  -2\tilde\psi_m \\
    &\tilde{\mathbb{Q}} \cdot \psi_m = -2i[V_m,C] \\
    &\tilde{\mathbb{Q}} \cdot \tilde\psi_m =   -\sqrt{2}\, \eD_m C \\
    &\tilde{\mathbb{Q}} \cdot \chi^+_{mn} =  2\sqrt{2}\,(\eD_{[m}
    V_{n]})^+ \\
    &\tilde{\mathbb{Q}} \cdot \chi^-_{mn} =  F^-_{mn} - 2i[V_m,V_n]^- \\
    &\tilde{\mathbb{Q}} \cdot \eta =  -2\sqrt{2}\,(\eD_m V^m +2\mu B) \\
    &\tilde{\mathbb{Q}} \cdot\tilde\eta =  -2i[B,C] \\
    &\tilde{\mathbb{Q}} \cdot B =  -\sqrt{2}\, \tilde\eta \\
    &\tilde{\mathbb{Q}} \cdot C =  0 \\
    &\tilde{\mathbb{Q}} \cdot V_m =  -\sqrt{2}\, \psi_m~,
  \end{aligned}
\end{equation}
the only change in comparison with the transformations on the original
fields \eqref{eq:BRST} being the explicit $\mu$-dependent terms in
$\mathbb{Q} \cdot \tilde\eta$ and $\tilde{\mathbb{Q}} \cdot \eta$.

The field redefinition \eqref{eq:redefs} takes the action
\eqref{eq:S0Lozano} to
\begin{equation}
 \label{eq:S1action}
 \begin{aligned}[t]
 \mathcal{S}^{(1)} = & 2\pi \tau k + \frac{1}{ e^2}\int d^4 x\,
  \tr\, \biggl( - \eD_m B\eD^m C -\eD_m V_n\eD^m V^n -\half F^+_{mn}
    F^{+mn} \\
 & + \eD_m\psi_n (4\chi^{+ mn} - \delta^{mn} \eta) + \eD_m\tilde\psi_n
    (4\chi^{- mn} - \delta^{mn} \tilde\eta)\\
 & - \tfrac{i}{8\sqrt{2}}\,((4\chi^+_{mn} -\delta_{mn}
    \eta)[4\chi^{+mn} -\delta^{mn} \eta,C] + ( 4\chi^-_{mn}
    -\delta_{mn} \tilde\eta)[4\chi^{-mn}  -\delta^{mn} \tilde\eta,C])
    \\
&  - i \sqrt{2}\,((4\chi^+_{mn} -\delta_{mn} \eta)[\tilde\psi^{m},
    V^{n}] - (4\chi^-_{mn} - \delta_{mn} \tilde\eta)[\psi^{m} ,
    V^{n}])\\
&  + i \sqrt{2}\,(\psi_m[\psi^m,B] + \tilde\psi_m[\tilde\psi^m,B])
  -\half[B,C]^2 +2[B,V_m][C,V^m]\\
&  + [V_m,V_n] [V^m,V^n] + 4\mu V^m \eD_m B - 2\mu\tilde\eta \eta -
    4\mu^2 B^2 \biggr)~,
\end{aligned}
\end{equation}
where $k$ is the instanton number
\begin{equation}
  k :=  \frac{1}{32\pi^2} \int d^4x\, \tr *F_{mn}F^{mn} ~.
\end{equation}
The redefinitions produce the additional terms proportional to $\mu$
and $\mu ^2$ in the above action; the explicit $x$-dependence in the
redefinitions \eqref{eq:redefs} drops out from the action, and hence
it is manifestly invariant under the translations.  The modified
action \eqref{eq:S1action} is invariant under the standard
$\PP,\MM,\DD$ transformations \eqref{eq:newPs}, \eqref{eq:newMDs}, the
modified conformal boosts, $\KK$ \eqref{eq:newKs}, and the CBRST
transformations \eqref{eq:newCBRSTs}.  The generators $\PP,\MM,\DD,
\KK$ satisfy the $\SO(5,1)$ algebra \eqref{eq:so51algebra}.  The CBRST
transformations square to zero, anticommute with each other and
commute with these $\SO(5,1)$ symmetries.

The change in the $F_{mn}$ terms from \eqref{eq:S0Lozano} to
\eqref{eq:S1action} is obtained by decomposing $F_{mn}$ into its
self-dual and anti self-dual parts, then recasting the topological
instanton number term (with parameter $\theta$) in terms of the
modular parameter $\tau$ defined in \eqref{eq:tau}.  The action
\eqref{eq:S1action} can then be conveniently expressed (using the
$\chi^{\pm}_{mn},\eta$ and $\tilde\eta$ equations of motion) as a
CBRST exact term (or an anti-CBRST exact term), plus a term depending
only upon $\tau$ and the instanton number $k$:
\begin{equation}
  \mathcal{S}^{(1)} =  \mathbb{Q} \cdot \Psi^{(1)} + 2\pi \tau k = 
  \tilde {\mathbb{Q}} \cdot \tilde\Psi^{(1)} - 2\pi \bar{\tau} k~,   
\end{equation}
where
\begin{equation}
 \label{eq:CBRSTexact}
 \begin{aligned}[t]
  \Psi^{(1)} =  \frac{1}{e^{2}_{0}} \int d^4 x\, \tr \biggl(& \half
    (F^+_{mn} -2i[V_m,V_n]^+)\chi^{+mn} - \sqrt{2}
    (\eD_{[m}V_{n]})^- \chi^{-mn} + \tfrac{1}{\sqrt{2}} \mu B
    \tilde\eta \\
   & - \tfrac{1}{2\sqrt{2}} V^m \eD_m \tilde\eta +
     \tfrac{1}{\sqrt{2}} B (\eD_m \psi^m + i \sqrt{2}\,[\tilde\psi_m ,
     V^m]) + \tfrac{i}{4} \eta [B,C] \biggr)~,
\end{aligned}
\end{equation}
and
\begin{equation}
 \label{eq:antiCBRSTexact}
\begin{aligned} 
\tilde\Psi^{(1)} =  \frac{1}{e^{2}_{0}} \int d^4 x\, \tr \biggl(
  &  - \half (F^-_{mn} - 2 i [V_m,V_n]^-)\chi^{-mn} -\sqrt{2}
    (\eD_{[m}V_{n]})^+\chi^{+mn} + \tfrac{1}{\sqrt{2}} \mu B \eta \\
   & - \tfrac{1}{2\sqrt{2}} V^m \eD_m \eta - \tfrac{1}{\sqrt{2}} B
      (\eD_m \tilde\psi^m -i\sqrt{2}\,[\psi_m , V^m]) - \tfrac{i}{4}
      \tilde\eta [B,C] \biggr)~.
\end{aligned}
\end{equation}
The topological observables will be discussed in section 6.


\section{Conserved currents}
\label{sec:currents}

We have seen that the action for the twisted theory in $\RR^4$ is
CBRST exact (modulo the instanton number term). We will now discuss
the conformal currents of the theory, and the derivation of the
appropriate flat-space energy-momentum tensor.

We begin by calculating the canonical energy-momentum tensor
$T^{\text{c}}_{mn}$, defined in general by
\begin{equation}
\label{eq:Tcanonical}
  T^{\text{c}}_{mn} :=  \tr \left(\sum_\Phi \d_n \Phi \cdot \Pi_m (\Phi)
  -\delta_{mn} \mathcal{L} \right) \qquad\text{where}\qquad
  \Pi_m(\Phi) :=  {\d \mathcal{S}}/{\d(\d^m \Phi)}
\end{equation}
for an action of the form $\mathcal{S} = \int d^4 x \tr \mathcal{L}$
with Lagrangian $\mathcal{L}(\Phi, \eD_m \Phi)$ and some set of
fields $\{\Phi\}$. By construction, this
tensor is conserved (on-shell), $\d^m T^{\text{c}}_{mn} = 0$.
Noether's theorem implies that for every symmetry of the theory there
exists a conserved current.  For conformal symmetries, these conserved
currents are suitable moments of an improved energy momentum tensor.
This tensor is obtained by adding \emph{improvement terms} to
$T^{\text{c}}_{mn}$. A discussion of this can be found in
\cite{ColJac}.  The purpose of the first such improvement term is to
symmetrise the energy-momentum tensor.  This first additional term is
given by $\d^p X_{pmn}$, where
\begin{equation}
\label{eq:Belextra}
  X_{pmn} :=  \half \sum_{\Phi} \tr \left(
  \left(\Sigma_{mn}\Phi\right)\cdot\Pi_p - 
    \left(\Sigma_{pm}\Phi\right)\cdot\Pi_n -
    \left(\Sigma_{pn}\Phi\right)\cdot\Pi_m \right),
\end{equation}  
and $\Sigma_{mn}$ are the usual spin generators. The expression $\d^p
X_{pmn}$ is automatically conserved as a result of the $(m
\leftrightarrow p)$ antisymmetry in $X_{pmn}$, and when added to
$T^{\text{c}}_{mn}$ constructs what is called the {\it Belinfante
  tensor}
\begin{equation}
\label{eq:Beldef}
T^{\text{B}}_{mn} \equiv T^{\text{c}}_{mn} + \d^p X_{pmn},
\end{equation}
which is symmetric and conserved on-shell. Evaluating this object for
the twisted theory \eqref{eq:S1action} gives
\begin{equation}
\label{eq:BelT}
  \begin{split}  
    T^{\text{B}}_{mn} =  \tr \biggl( &-2\eD_{(m}B\eD_{n)}C -2\eD_m V_p
    \eD_n V^p -F_{mp}F_{n}^{p} \\ &+2\eD_p [(\eD_{(m} V^p) V_{n)}
    -(\eD_{(m} V_{n)})V^p ] + 4\mu V_{(m}\eD_{n)} B \\
    &-8 \chi^{+\;p}_{m} (\eD_{[n} \psi_{p]})^- - 4 i
    \sqrt{2}\,([\chi^+_{p(m} , V_{n)}]\tilde\psi^p + 2\chi^{+\;p}_{m}
    [V_{[n} , \tilde\psi_{p]}]^-) \\
    & -8 \chi^{-\;p}_{m} (\eD_{[n} \tilde\psi_{p]})^+ + 4 i
    \sqrt{2}\,([\chi^-_{p(m}, V_{n)}]\psi^p +2\chi^{-\;p}_{m} [V_{[n}
    , \psi_{p]}]^+) \\
    & - 2(\eD_{(m}\eta)\psi_{n)} - i
    \sqrt{2}\,[\eta,V_{(m}]\tilde\psi_{n)} -2(\eD_{(m}\tilde\eta)
    \tilde\psi_{n)} + i \sqrt{2}\,[\tilde\eta,V_{(m}]\psi_{n)} \\
    & + 4 i \sqrt{2}\,\chi^{+\;p}_{m} [\chi^+_{np},C]
    +2i\sqrt{2}\,\psi_m [\psi_n ,B] +\delta_{mn} \eD_p (\eta \psi^p) \\
    & + 4 i \sqrt{2}\,\chi^{-\;p}_{m} [\chi^-_{np},C] + 2 i
    \sqrt{2}\,\tilde\psi_m [\tilde\psi_n ,B] +\delta_{mn} \eD_p 
(\tilde\eta
    \tilde\psi^p) -\delta_{mn} {\mathcal{L}}^{(1)} \biggr)~.
  \end{split}
\end{equation}

The second improvement term is specific for conformally symmetric
theories and requires the calculation of the \emph{field-virial}
$\mathcal{V}_m$, defined as
\begin{equation}
\label{eq:virial}
  {\mathcal{V}}_m \equiv \tr \left( \left(\delta_{mn} \Delta_\Phi +
  \Sigma_{mn}\right)\Phi\;\cdot\Pi^n \right)~,
\end{equation}
where $\Delta_\Phi$ is the conformal weight of $\Phi$.  
It can then be shown that the
condition of special conformal invariance requires that
\begin{equation}
\label{eq:virial2}
\mathcal{V}_m =  \d^n \sigma _{mn},
\end{equation}
for some function of the fields $\sigma_{mn}$. 

For the twisted theory \eqref{eq:S1action} this condition is
satisfied, with
\begin{equation}
\label{eq:sigmavirial}
\begin{aligned}
\sigma_{mn} = &\tr \bigl(-2g_{mn}BC -2V_m V_n -\delta_{mn}V^2 
+\mu^{-1}
    (V_m \eD_n C - (\eD_n V_m) C) \\
    &\qquad\qquad {} -2\mu^{-1} \tilde\psi_{(m}\psi_{n)}
    +\mu^{-1} \delta_{mn} \tilde\psi_p \psi^p -\mu^{-1}
    \epsilon_{mnpq} \tilde\psi^p \psi^q \bigr),
\end{aligned}
\end{equation}
and hence
\begin{equation}
\label{eq:virial3}
  \mathcal{V}_m =  -2B(\eD_m C) +8\mu V_m B +2(\eD_n
    V^n)V_m -4(\eD_{(m}V_{n)})V^n -2\eta \psi_m -2\tilde\eta
    \tilde\psi_m,  
\end{equation}
where we used the twisted conformal weights in this calculation.
Next, the prescription is to take the symmetric object 
\begin{equation}
\label{eq:symm}
         s_{mn}:= \sigma_{(mn)},
\end{equation}
  and use it to construct
\begin{equation}
\label{eq:Ydef}
  Y_{pqmn} =  \delta_{pq}s_{mn} + \delta_{mn}s_{pq} -
  \delta_{pm}s_{qn} -\delta_{pn}s_{qm} - \tfrac{1}{3}
  (\delta_{pq}\delta_{mn} - \delta_{pm}\delta_{qn})s^{k}_{k}.
\end{equation}
The second addition to the energy-momentum tensor is then $\half \d^p
\d^q Y_{pqmn}$, which on its own is both symmetric and conserved.
Thus, for flat-space theories with full conformal invariance, the
final improved form of the energy-momentum tensor $T_{mn}$ is given by
\begin{equation}
\label{eq:Tfinal}
T_{mn} := T^{\text{B}}_{mn} + \half \d^p \d^q Y_{pqmn}.
\end{equation}
This fully improved energy-momentum tensor $T_{mn}$ is also traceless,
as required in order to construct conformal currents as the moments
$J_{im} = k_i^n T_{mn}$, using the usual 15 conformal Killing vectors
$k_i^n$ ($i=1,...,15$) in flat space.

For the twisted theory with action \eqref{eq:S1action}, this second
improvement term is explicitly given by
\begin{equation}
  \label{eq:Ytwisted}
  \begin{split}
    \half \d^p \d^q Y_{pqmn} = \tr &\biggl( (\delta_{mn}\eD^2
    -\eD_{(m} \eD_{n)}) [-\tfrac23 BC -\tfrac16 \mu^{-1}(V^k \eD_k C
    -(\eD_k V^k) C) + \tfrac23 \mu^{-1} \tilde\psi^k \psi_k ] \\
    &+ \eD^2 [ -V_m V_n +\half \mu^{-1} (V_{(m}\eD_{n)} C -
    (\eD_{(m}V_{n)})C) - \mu^{-1} \tilde\psi_{(m} \psi_{n)} ] \\
    &+ \delta_{mn}\eD^p \eD^q [ -V_p V_q +\half \mu^{-1}
    (V_{(p}\eD_{q)} C - (\eD_{(p}V_{q)})C) -\mu^{-1} \tilde\psi_{(p}
    \psi_{q)} ] \\
    &-\eD_m \eD^p [ -V_n V_p +\half \mu^{-1} (V_{(n}\eD_{p)} C -
    (\eD_{(n}V_{p)})C) -\mu^{-1} \tilde\psi_{(n} \psi_{p)} ] \\
    &-\eD_n \eD^p [ -V_m V_p +\half \mu^{-1} (V_{(m}\eD_{p)} C -
    (\eD_{(m}V_{p)})C) -\mu^{-1} \tilde\psi_{(m} \psi_{p)} ] \biggr).
  \end{split}
\end{equation}
This expression is to be added to the expression in \eqref{eq:BelT} to
give the final traceless, symmetric, conserved energy-momentum tensor
$T_{mn} := T^{\text{B}}_{mn} + \half \d^p \d^q Y_{pqmn}$ for the
theory \eqref{eq:S1action}.  This improved energy-momentum tensor
$T_{mn}$ is CBRST exact, with $T_{mn} = \mathbb{Q}\cdot G_{mn}$ for a
certain function $G_{mn}$ of the fields.  This is given by the flat
space specialisation of the result in the next section.


\section{Coupling to gravity}
\label{sec:gravity}

We now turn to the formulation of the topological theory on a
Riemannian 4-manifold $\mathcal{M}$ with metric $g_{mn}$, and seek a
BRST-invariant action which is invariant under Weyl scalings and which
reduces to the flat space theory considered in previous sections.  The
energy-momentum tensor will be defined as usual in terms of the action
$\mathcal{S}$ by
\begin{equation}
\label{eq:Enmom}
  T_{mn} =  \frac{2}{\sqrt{g}} \frac{\delta
    {\mathcal{S}}}{\delta g^{mn}}.
\end{equation}

The first step to coupling a flat-space gauge invariant theory to a
general curved background is with a minimal prescription of
covariantising the action via $\delta_{mn} \mapsto g_{mn}$, $d^4 x
\mapsto \sqrt{g}\,d^4 x$ and $\eD \mapsto \nabla \equiv \eD + \Gamma$
(where $\eD$ is the gauge covariant derivative and $\Gamma (g)$ is the
Levi-Civita connection on $\mathcal{M}$). In general, if this is done
the resulting energy-momentum tensor, as defined by \eqref{eq:Enmom},
will reduce in flat space to the Belinfante tensor \eqref{eq:Beldef}.
(Note that fields satisfying a metric-dependent constraint, such as
the fields $\chi^{\pm}_{mn}$ when defined on a curved manifold, will
also transform under metric variations (see \cite{WittenTFT}).)  For
conformal theories in flat space one needs to add a second improvement term in
the energy-momentum tensor, giving  \eqref{eq:Tfinal}. The
corresponding curved space theory must include additional non-minimal
terms which reproduce this second improvement term $ \half \d^p \d^q
Y_{pqmn}$ in the flat-space case. The extra terms required are
\begin{equation}
\label{eq:extraterms}
   \mathcal{S}^{\partial\partial Y} :=   \int_{\mathcal{M}} d^4 x
    \sqrt{g}\, \left( \half R_{mn} s^{mn} - \frac{1}{12} R
      s^{k}_{k}\right),   
\end{equation}
where $R_{mn}$ and $R$ are  the Ricci tensor and scalar of
${\mathcal{M}}$ while $s_{mn}$ is the minimally coupled version of the
expression given by  \eqref{eq:symm} and \eqref{eq:sigmavirial} in the
previous section.

However, in the case of the theory with action \eqref{eq:S1action}, it
turns out that defining such a theory in curved space via minimal
coupling and adding the above terms does {\it not} produce a
topological conformal field theory---for example, the resulting
energy-momentum tensor defined by \eqref{eq:Enmom} is not traceless
or CBRST exact in curved space.  One is, however, free to add further
curvature dependent terms to this action in order to obtain if
possible a curved space theory which has the required properties and
 reduces in flat space to the theory given in Section 4.  It does 
turn out to be possible to find such additional terms,
and this curved space conformal topological quantum field theory will
now be presented and discussed.

The action for this conformal TQFT is given by
\begin{equation}
\label{eq:finalaction}
  \mathcal{S} =   \mathcal{S}^{m.c.} +  \mathcal{S}^{\partial\partial Y}
                   +  \mathcal{S}^{CR} + \mathcal{S}^{\text{Euler}}, 
\end{equation}
where $\mathcal{S}^{m.c.}$ is the minimally-coupled version of the
action \eqref{eq:S1action}, $\mathcal{S}^{\partial\partial Y}$ is the
action \eqref{eq:extraterms} above, and the other two terms on the
right-hand side of \eqref{eq:finalaction} are given by
\begin{equation}
  \label{eq:R-s}
  \mathcal{S}^{CR} =  -\frac{1}{144\mu^2e^{2}}\int_{\mathcal{M}}
  d^4 x  \sqrt{g}\, \tr C^2 R^2~
\end{equation}
and
\begin{equation}
\label{eq:Euleraction}
  \mathcal{S}^{\text{Euler}} =  \frac{\pi^2}{e^2 \mu^2} 
               \int_{\mathcal{M}} d^4 x\, \sqrt{g}\, e(\mathcal{M}) \tr C^2
\end{equation}
where
\begin{equation}
\label{eq:Eulerdensity}  
e(\mathcal{M}) := \frac{1}{32 \pi^2}\bigl(W_{mnpq} W^{mnpq} -2 R_{mn}
  R^{mn} + \tfrac{2}{3} R^2\bigr)
\end{equation}
is the Euler density on $\mathcal{M}$.

The action \eqref{eq:finalaction} has the following properties:
\begin{itemize}
\item $\mathcal{S}$ is invariant under nilpotent CBRST transformations
  $\mathbb{Q}$ and $\tilde{\mathbb{Q}}$, with
  $\{\mathbb{Q},\tilde{\mathbb{Q}}\}=0$.
\item $\mathcal{S}$ is CBRST and anti-CBRST exact.
\item The energy-momentum tensor $T_{mn}$ arising from $\mathcal{S}$
  is CBRST and anti-CBRST exact.
\item $T_{mn}$ is traceless, and the corresponding local Weyl
  symmetries of $\mathcal{S}$ commute with the CBRST symmetries.
\end{itemize}
\noindent To see these properties, begin by rewriting the action
$\mathcal{S}$ of \eqref{eq:finalaction} in the form
\begin{equation}
\label{eq:Saction}
  \mathcal{S} =   \mathcal{S}^{(2)} + \mathcal{S}^{\text{Euler}}, 
\end{equation}
with
\begin{equation}
\label{eq:S2action}
  \begin{split}
    \mathcal{S}^{(2)} &=  \frac{1}{e^2}\int_{\mathcal{M}} d^4 x
    \sqrt{g} \tr \biggl( -\half (-F^+_{mn} + 2i [V_m , V_n]^+ )^2
    -4((\nabla_{[m}V_{n]})^-)^2-\nabla_m B\nabla^m C \\
    &-(\nabla_m
    V^m +2\mu B +\frac{1}{12\mu} CR)^2 +\frac{1}{\mu} (V^m \nabla^n C
    -\tilde\psi^m \psi^n ) (R_{mn} - \frac{1}{3} g_{mn}R) \\
    &+\nabla_m\psi_n (4\chi^{+ mn} - g^{mn} \eta) +
    \nabla_m\tilde\psi_n
    (4\chi^{- mn} - g^{mn} \tilde\eta) -2\mu\tilde\eta \eta \\
    &-\frac{i}{8\sqrt{2}}\,((4\chi^+_{mn} -g_{mn} \eta)[4\chi^{+mn}
    -g^{mn} \eta,C] +(4\chi^-_{mn} -g_{mn} \tilde\eta)[4\chi^{-mn}
    -g^{mn} \tilde\eta,C]) \\
    &-i\sqrt{2}\,((4\chi^+_{mn} -g_{mn}
    \eta)[\tilde\psi^{m} , V^{n}] - (4\chi^-_{mn} -g_{mn}
    \tilde\eta)[\psi^{m} , V^{n}]) \\
    &+i\sqrt{2}\,(\psi_m[\psi^m,B] +
    \tilde\psi_m[\tilde\psi^m,B]) -\half[B,C]^2 +2[B,V_m][C,V^m]
    \biggr) + 2\pi \tau k~,
  \end{split}
\end{equation}
and $\mathcal{S}^{\text{Euler}}$ given in \eqref{eq:Euleraction}.

The curved space CBRST symmetries of the action \eqref{eq:Saction}
are given by ($\nabla_m$ is the gauge and diffeomorphism covariant
derivative)
\begin{equation}
\label{eq:curvedCBRSTs}
  \begin{aligned}[t]
    &\mathbb{Q} \cdot A_m =  2\psi_m\\
    &\mathbb{Q} \cdot \psi_m =  \sqrt{2}\, \nabla_m C\\
    &\mathbb{Q} \cdot \tilde\psi_m = -2i[V_m,C]\\
    &\mathbb{Q} \cdot \chi^+_{mn} =  -F^+_{mn} +2i[V_m,V_n]^+\\
    &\mathbb{Q} \cdot \chi^-_{mn} =  2\sqrt{2}\,(\nabla_{[m}
    V_{n]})^-\\
    &\mathbb{Q} \cdot \eta =  2i[B,C]\\
    &\mathbb{Q} \cdot \tilde\eta =  -2\sqrt{2}(\nabla_m V^m +2\mu B +
    \tfrac{1}{12\mu} CR)\\
    &\mathbb{Q} \cdot B =  \sqrt{2}\, \eta\\
    &\mathbb{Q} \cdot C =  0\\
    &\mathbb{Q} \cdot V_m =  -\sqrt{2}\, \tilde\psi_m\\
    &\mathbb{Q} \cdot g_{mn} =  0
  \end{aligned}
  \qquad
  \begin{aligned}[t]
    &\tilde{\mathbb{Q}} \cdot A_m =  -2\tilde\psi_m \\
    &\tilde{\mathbb{Q}} \cdot \psi_m = -2i[V_m,C] \\
    &\tilde{\mathbb{Q}} \cdot \tilde\psi_m =  -\sqrt{2}\, \nabla_m C \\
    &\tilde{\mathbb{Q}} \cdot \chi^+_{mn} =  2\sqrt{2}\,(\nabla_{[m}
    V_{n]})^+ \\
    &\tilde{\mathbb{Q}} \cdot \chi^-_{mn} =  F^-_{mn} - 2i[V_m,V_n]^- \\
    &\tilde{\mathbb{Q}} \cdot \eta =  -2\sqrt{2} (\nabla_m V^m +2\mu B
    +\tfrac{1}{12\mu} CR) \\
    &\tilde{\mathbb{Q}} \cdot\tilde\eta =  -2i[B,C] \\
    &\tilde{\mathbb{Q}} \cdot B =  -\sqrt{2}\, \tilde\eta \\
    &\tilde{\mathbb{Q}} \cdot C =  0 \\
    &\tilde{\mathbb{Q}} \cdot V_m =  -\sqrt{2}\, \psi_m\\
    &\tilde{\mathbb{Q}} \cdot g_{mn} =  0  ~.
  \end{aligned}
\end{equation}
The terms $ \mathcal{S}^{(2)} $ and $ \mathcal{S}^{\text{Euler}} $ are
separately invariant.  These CBRST transformations differ from the
minimally-coupled version of the flat-space transformations
\eqref{eq:newCBRSTs} by the addition of $CR$ terms to $\mathbb{Q}
\cdot \tilde\eta$ and $\tilde{\mathbb{Q}} \cdot \eta$. Since $C$ and
the metric $g_{mn}$ are invariant under $\mathbb{Q}$ and
$\tilde{\mathbb{Q}}$, these terms do not affect the calculation of
anticommutators of the CBRST transformations, and indeed one can
readily show that $\mathbb{Q}$ and $\tilde{\mathbb{Q}}$ square to zero
and anti-commute with each other, up to a gauge transformation and
using the equations of motion following from the action
\eqref{eq:finalaction}.

The action \eqref{eq:Saction} is not only CBRST and anti-CBRST
invariant, but is also exact. To see this, first note that
${\mathcal{S}}^{(2)}$ can be written in the CBRST exact forms:
\begin{equation}
 \label{eq:S2exact}
 \begin{split}
    \mathcal{S}^{(2)} &=  \mathbb{Q} \cdot\Psi^{(2)} + 2\pi \tau k +
    \text{on-shell terms} \\
    &=  \tilde{\mathbb{Q}} \cdot \tilde\Psi^{(2)} - 2\pi \bar\tau k +
    \text{on-shell terms},
  \end{split}
\end{equation}
where
\begin{equation}
\label{eq:psiform} 
 \begin{split}
    \Psi^{(2)} =  \frac{1}{e^{2}} \int_{\mathcal{M}} d^4 x
    \sqrt{g} & \tr \biggl( \half (F^+_{mn}
    -2i[V_m,V_n]^+)\chi^{+mn}
    -\sqrt{2}\,(\nabla_{[m}V_{n]})^-\chi^{-mn} \\
    &+\tfrac{1}{2\sqrt{2}} ( \nabla_m V^m +2\mu B +\tfrac{1}{12\mu} CR
    )\tilde\eta +\tfrac{1}{\sqrt{2}\,\mu} V^m \psi^n ( R_{mn} -
    \tfrac{1}{3} g_{mn}R) \\
    &+\tfrac{1}{\sqrt{2}} B (\nabla_m \psi^m
    +i\sqrt{2}\,[\tilde\psi_m , V^m]) +\tfrac{i}{4} \eta [B,C]
    \biggr)
\end{split}
\end{equation}
and
\begin{equation}
\label{eq:psis}
  \begin{split}
    \tilde\Psi^{(2)} =  \frac{1}{e^{2}} \int_{\mathcal{M}} d^4 x
    \sqrt{g}&\tr \biggl( -\half (F^-_{mn} -2i[V_m,V_n]^-)
    \chi^{-mn} -\sqrt{2}\,(\nabla_{[m}V_{n]})^+\chi^{+mn} \\  
    &+\tfrac{1} {2\sqrt{2}} (\nabla_m V^m +2\mu B +\tfrac{1}{12\mu}CR)
    \eta -\tfrac{1}{\sqrt{2}\,\mu} V^m \tilde\psi^n (R_{mn}
    -\tfrac{1}{3}g_{mn}R) \\
    &-\tfrac{1} {\sqrt{2}} B (\nabla_m \tilde\psi^m
    -i\sqrt{2}\,[\psi_m, V^m]) -\tfrac{i}{4} \tilde\eta [B,C]
    \biggr)~.
\end{split}
\end{equation}
Furthermore, locally the Euler density is a total derivative and can
be written as the divergence of some $Z_m$:
\begin{equation}
e(\mathcal{M}) = \frac{1}{32 \pi^2}\bigl(W_{mnpq} W^{mnpq} -2 R_{mn}
  R^{mn} + \tfrac{2}{3} R^2\bigr) = \nabla^mZ_m.
\end{equation}
%
It is then easy to see that the 
Lagrangian $\mathcal{L}^{\text{Euler}}$ for the 
action $\mathcal{S}^{\text{Euler}}=  \int_{\mathcal{M}} d^4 x\, \sqrt{g}
\mathcal{L}^{\text{Euler}}
$ is
also exact up to derivative terms:
\begin{equation}
  \label{eq:Eulerexact}
  \mathcal{L}^{\text{Euler}} =  \mathbb{Q}\Lambda  +
  \mathbb{V}=  \tilde{\mathbb{Q}}\tilde{\Lambda}  +
  \mathbb{V},
\end{equation}
where
\begin{equation}
  \label{eq:Lambdadefs}
  \begin{aligned}
    \Lambda & := -\sqrt{2}
    \frac{\pi^2}{e^2\mu^2}
    \, Z^m \tr (C\psi_m), \\
    \tilde{\Lambda} & := \sqrt{2}
    \frac{\pi^2}{e^2\mu^2}
   \, Z^m \tr (C\tilde\psi_m),
  \end{aligned}
\end{equation}
and
\begin{equation}
  \label{Zdef}
  \mathbb{V}:= 
  \frac{\pi^2}{e^2 \mu^2} 
  \, \nabla _ m\left(  Z^m \tr C^2\right).
\end{equation}
Although $Z_m$ is not globally well-defined, any two choices of $Z_m$
will differ by a globally well-defined vector field, and the change in
$Z_m$ under a diffeomorphism will be similarly well-defined, so that
the ambiguity in the action $\mathcal{S}^{\text{Euler}}$ will be the
sum of a surface term (which will vanish for compact ${\mathcal{M}}$
or with suitable boundary conditions) and an exact term (which will
not contribute to the functional integral).

The next step is to show exactness of the energy-momentum tensor of
the theory defined by the action \eqref{eq:finalaction}, with CBRST
symmetries \eqref{eq:curvedCBRSTs}. First consider the action
$\mathcal{S}^{(2)}$ of \eqref{eq:S2action} (a similar argument to the
following appears in a related context in \cite{Spence-BI}. See also
\cite{Vandoren-BV} for a discussion of topological field theory in the BV
formalism, which includes an analysis of metric dependence in the BRST
transformations). Varying \eqref{eq:S2action} with respect to the metric we
find \begin{equation} \label{eq:splitit}   \begin{split}
    \delta_g \mathcal{S}^{(2)} &= \mathbb{Q} \cdot (\delta_g
    \Psi^{(2)}) +  [\delta_g , \mathbb{Q}]\Psi^{(2)} +
    \delta_g \{\text{on-shell terms}\}  \\
    &=  \mathbb{Q} \cdot \biggl( \delta_g \Psi^{(2)} +
          \frac{1}{2e^{2}\sqrt{2}}
    \int_{\mathcal{M}} d^4 x \sqrt{g}\,\tr \delta_g (\nabla_m V^m) \tilde\eta
    \biggr) + \text{(on-shell terms).}
\end{split}
\end{equation}
(The $\delta_g \{\text{on-shell terms}\}$ in the first line in
\eqref{eq:splitit} involve the $\tilde\eta$ and $\chi^\pm$ equations
of motion. The metric variations of the $\chi^\pm$ equations of motion
only generate further equations of motion terms. The metric variation
of the $\tilde\eta$ equation of motion term however generates a term
which does not vanish on-shell. However, this term combines with a
second term to give the CBRST exact term involving $\nabla_m V^m$ in
the second line in \eqref{eq:splitit}. This second term arises from
the commutator $[\delta_g , \mathbb{Q}]$, which only acts
non-trivially on the $\tilde\eta$ terms in $\Psi^{(2)}$.
Alternatively, one can work with the form of the action given below
for which the CBRST transformations are nilpotent off-shell, and show
that the metric dependence is CBRST exact without using
equations of motion.)  Now, writing %
\begin{equation}
  \delta_g \Psi^{(2)} +  \frac{1}{2e^{2}\sqrt{2}} \int_{\mathcal{M}} d^4 x
  \sqrt{g}\,\tr  \delta_g (\nabla_m V^m)
  \tilde\eta  = 
  \frac{1}{2e^{2}}\int_{\mathcal{M}} d^4 x  \sqrt{g}\,\delta  
  g^{mn} G^{(2)}_{mn},
\end{equation}
for some $G^{(2)}_{mn}$ defined by the above relation, it follows that
the energy-momentum tensor derived from the action $\mathcal{S}^{(2)}$
is CBRST exact with
\begin{equation}
  T^{(2)}_{mn} =  \mathbb{Q} \cdot G^{(2)}_{mn} + \text{on-shell 
terms,}
\end{equation}
with the analogous statements holding for the $\ZZ_2$ related
generator $\tilde{\mathbb{Q}}$.
The other contribution to the full energy-momentum tensor $T_{mn}$ of
the full action \eqref{eq:Saction} comes from the action
\eqref{eq:Euleraction}.  Since this action is exact (up to the
topological term involving $\mathbb{V}$), and only contains the
CBRST inert fields $C, g_{mn}$, it follows directly that the metric
variation of this action is also exact.  Thus the full energy-momentum
tensor derived from the action \eqref{eq:finalaction} is CBRST (and
anti-CBRST) exact.

The final property of the theory defined by \eqref{eq:finalaction} is
that the trace of the energy-momentum tensor vanishes on-shell.  A
straightforward calculation based upon the action \eqref{eq:S2action}
yields the result that the trace of the corresponding energy-momentum
tensor is given by
\begin{equation}
 g^{mn} T^{(2)}_{mn} =  \tfrac{1}{4\mu^2} \nabla^m\nabla^n \left( (R_{mn}
  -\half g_{mn}R)\tr (C^2) \right)~.
\end{equation}
This is precisely cancelled by the trace of the energy-momentum tensor
coming from the other part \eqref{eq:Euleraction} of the full action.
Then, since the theory with action \eqref{eq:finalaction} has a
traceless energy-momentum tensor on-shell, this action is Weyl
invariant under a combination of local Weyl rescalings of the metric
$g_{mn} \mapsto \exp (-2w(x))\,g_{mn}$ and the action of the Weyl
symmetries on the fields.  The latter are given by
\begin{equation}
\label{eq:Weylsymmetries}
  \begin{aligned}[t]
    &\delta_W A_m =0 \\
    &\delta_W \psi_m = 0 \\
    &\delta_W \tilde\psi_m = 0 \\
    &\delta_W \chi^+_{mn} = \mu^{-1} (\tilde\psi_{[m}\nabla_{n]}w)^+
    \\
    &\delta_W \chi^-_{mn} = - \mu^{-1} (\psi_{[m}\nabla_{n]}w)^-
  \end{aligned}
  \qquad
  \begin{aligned}[t]
    &\delta_W \eta = 2w\eta - \mu^{-1} \tilde\psi_m \nabla^m w \\
    &\delta_W \tilde\eta = 2w\tilde\eta + \mu^{-1} \psi_m \nabla^m w \\
    &\delta_W B = 2wB + \mu^{-1} V_m \nabla^m w \\
    &\delta_W C = 0 \\
    &\delta_W V_m = - \half \mu^{-1} C\nabla_m w~.
  \end{aligned}
\end{equation}
These curved-space Weyl transformations contain non-conventional
$1/\mu$ terms, as one would expect from the presence of such terms in
the flat-space conformal transformations.  These Weyl symmetries
$\delta_W$ commute with the CBRST transformations $\mathbb{Q}$ and
$\tilde{\mathbb{Q}}$.

On-shell conditions appearing in the curved space formulation in this
section can be lifted by writing off-shell formulations using
auxiliary fields, just as in the SO(4) twisted theory (see
\cite{Lozano}). The auxiliary fields required may be denoted $P,
N^{\pm}_{mn}$. The $\ZZ_2$ symmetry acts on these as $P\rightarrow -P,
N^\pm_{mn}\rightarrow -N^\mp_{mn}$.  The off-shell form of the action
\eqref{eq:S2action} is
\begin{equation}
\label{eq:S2actionoffshell}
  \begin{aligned}[t]
    \mathcal{S}^{(2)} &=  \frac{1}{e^2}\int_{\mathcal{M}} d^4 x
    \sqrt{g} \tr \biggl( 
    \frac{1}{2}N^+_{mn}\bigl(N^{mn} + 2F^{mn} -4i[V^m,V^n]\bigr) \\
  & + \frac{1}{2}N^-_{mn}\bigl(N^{mn} -4\sqrt{2}\nabla^{[m}V^{n]}\bigr) 
   + \frac{1}{8}P\bigl(P + 4\sqrt{2}(\nabla_mV^m+2\mu B+\frac{1}{12\mu}CR)\bigr)\\
&+\frac{1}{\mu} (V^m \nabla^n C
    -\tilde\psi^m \psi^n ) (R_{mn} - \frac{1}{3} g_{mn}R) -\nabla_m B\nabla^m C \\
    &+\nabla_m\psi_n (4\chi^{+ mn} - g^{mn} \eta) +
    \nabla_m\tilde\psi_n
    (4\chi^{- mn} - g^{mn} \tilde\eta) -2\mu\tilde\eta \eta \\
    &-\frac{i}{8\sqrt{2}}\,((4\chi^+_{mn} -g_{mn} \eta)[4\chi^{+mn}
    -g^{mn} \eta,C] +(4\chi^-_{mn} -g_{mn} \tilde\eta)[4\chi^{-mn}
    -g^{mn} \tilde\eta,C]) \\
    &-i\sqrt{2}\,((4\chi^+_{mn} -g_{mn}
    \eta)[\tilde\psi^{m} , V^{n}] - (4\chi^-_{mn} -g_{mn}
    \tilde\eta)[\psi^{m} , V^{n}]) \\
    &+i\sqrt{2}\,(\psi_m[\psi^m,B] +
    \tilde\psi_m[\tilde\psi^m,B]) -\half[B,C]^2 +2[B,V_m][C,V^m]
    \biggr) + 2\pi \tau k~.
  \end{aligned}
\end{equation}
The off-shell CBRST symmetries are given by
\begin{equation}
\label{eq:curvedCBRSTsoffshell}
  \begin{aligned}[t]
    &\mathbb{Q} \cdot A_m =  2\psi_m\\
    &\mathbb{Q} \cdot \psi_m =  \sqrt{2}\, \nabla_m C\\
    &\mathbb{Q} \cdot \tilde\psi_m = -2i[V_m,C]\\
    &\mathbb{Q} \cdot \chi^\pm_{mn} =  N^\pm_{mn}\\
    &\mathbb{Q} \cdot N^\pm_{mn} = 2i\sqrt{2}[\chi^\pm_{mn},C]\\
    &\mathbb{Q} \cdot \eta =  2i[B,C]\\
    &\mathbb{Q} \cdot \tilde\eta =  P\\
    &\mathbb{Q} \cdot P =  2i\sqrt{2}[\tilde\eta,C]\\
    &\mathbb{Q} \cdot B =  \sqrt{2}\, \eta\\
    &\mathbb{Q} \cdot C =  0\\
    &\mathbb{Q} \cdot V_m =  -\sqrt{2}\, \tilde\psi_m\\
    &\mathbb{Q} \cdot g_{mn} =  0
  \end{aligned}
  \qquad
  \begin{aligned}[t]
    &\tilde{\mathbb{Q}} \cdot A_m =  -2\tilde\psi_m \\
    &\tilde{\mathbb{Q}} \cdot \psi_m = -2i[V_m,C] \\
    &\tilde{\mathbb{Q}} \cdot \tilde\psi_m =  -\sqrt{2}\, \nabla_m C \\
    &\tilde{\mathbb{Q}} \cdot \chi^+_{mn} =  N^\pm_{mn} \\
       &\tilde{\mathbb{Q}} \cdot N^\pm_{mn} = 2i\sqrt{2}[\chi^\pm_{mn},C]\\  
&\tilde{\mathbb{Q}} \cdot \eta =  P \\
    &\tilde{\mathbb{Q}} \cdot\tilde\eta =  -2i[B,C] \\
&\tilde{\mathbb{Q}} \cdot P =  2i\sqrt{2}[\eta,C] \\
   &\tilde{\mathbb{Q}} \cdot B =  -\sqrt{2}\, \tilde\eta \\
    &\tilde{\mathbb{Q}} \cdot C =  0 \\
    &\tilde{\mathbb{Q}} \cdot V_m =  -\sqrt{2}\, \psi_m\\
    &\tilde{\mathbb{Q}} \cdot g_{mn} =  0  ~.
  \end{aligned}
\end{equation}
Then in this formulation with these
auxiliary fields
$\mathbb{Q}^2= \tilde{\mathbb{Q}}^2=0$ up to a gauge transformation  
with parameter $2\sqrt{2}iC$ 
and
$\{{\mathbb{Q}}, \tilde {\mathbb{Q}}\}=0$, without use of
field equations.  With these modifications, the action is given by the sum
of the topological term $2\pi \tau k$ plus an exact piece, without needing to
use field equations, and the metric variation of the action is also exact off-shell.   
However, the energy momentum  tensor is still only traceless on-shell.


Turning to the construction of observables for the conformal TQFT with
action \eqref{eq:finalaction}, we first need suitable operators
$\mathcal{O}$. The functionals studied in \cite{Marcus} are also
suitable here, and the following gauge invariant, CBRST closed (but
not exact) expressions arise:
\begin{equation}
\label{eq:invops}
  \begin{aligned}[t]
    {\mathcal{O}}_{0} &= \int_{\gamma_0}\tr C^2 \\
    {\mathcal{O}}_{1} &= \int_{\gamma_1} \sqrt{2} \tr\left( C\wedge\psi
    \right) \\
    {\mathcal{O}}_{2} &= \int_{\gamma_2} \half \tr\left( \psi\wedge\psi
    + \tfrac{1}{2\sqrt{2}} C\wedge F\right) \\
    {\mathcal{O}}_{3} &= \int_{\gamma_3} \tfrac14 \tr\left( \psi\wedge
    F\right) \\
    {\mathcal{O}}_{4} &= \int_{\gamma_4} \tfrac{1}{32} \tr \left(
    F\wedge F \right),  
  \end{aligned}
\end{equation}
together with a $\ZZ_2$ related set of observables corresponding to
the cohomology of $\tilde{\mathbb{Q}}$. The $\gamma_s$, $s=0,...,4$
are homology cycles on $\mathcal{M}$ ($\gamma_0$ is a point).  The
operators ${\mathcal{O}}_k$ and their anti-CBRST analogues satisfy the
descent equations given in \cite{Marcus}.  The integrands in
\eqref{eq:invops} are also invariant under the local scale
transformations \eqref{eq:Weylsymmetries}.  Note that the integrands
in any of these observables could be multiplied by any scalar function
of the curvature and its derivatives, such as $e(\mathcal{M})$, and
would remain CBRST closed but not exact.  However, these would only be
Weyl-invariant for special combinations of curvatures, such as the
square of the Weyl tensor.


\section{Theta Angle and S-Duality}
\label{sec:thet}

The Lorentzian theory has a real theta-term in the action
\begin{equation}
\label{eq:thett}
 S_r= -\frac{\theta}{ 32\pi^2}\int  \tr F \wedge F
\end{equation}
which is Wick rotated to an imaginary term
\begin{equation}
\label{eq:thetti}
 S_i= -i\frac{\theta}{ 32\pi^2}\int  \tr F \wedge F
\end{equation}
in the Euclideanised action.  As usual, the parameter $\theta$ must be
an angle, periodically identified, if the Euclideanised path integral
is to be single-valued, and the presence of the parameter $\theta$ leads
to theta-vacua instead of the naive vacuum.  However, the Euclidean
theory is written down directly in Euclidean space, and need have
nothing to do with any Wick rotation. In particular, if it is not
required to be the Euclideanisation of a real Lorentzian action, there
is no reason why the theta-term has to be imaginary, and one could
instead write down a real theta-term \eqref{eq:thett} in Euclidean
space. Indeed, it is natural to have a real action, and this also
follows from reduction from 9+1 dimensions \cite{Hulltym}.  To see
this, consider including a term
\begin{equation}
\label{eq:cett}
 S =  \int C\wedge \tr (F \wedge F)
\end{equation}
in the 9+1 dimensional Yang-Mills action, where $C$ is a background
6-form.  Such terms arise for example in considering D5-D9 brane
systems.  Then reducing on a Euclidean 6-torus gives, among other
terms, the real theta-term \eqref{eq:thett} in 3+1 dimensions, while
reducing on five space and one time dimension again gives a real
theta-term, but this time in Euclidean space. If theta is real in
Euclidean space, there is no reason to require theta to be an angle.

In the Lorentzian or the Euclideanised theory, the angle theta can be
combined with the coupling constant $e$ into a complex variable $\tau$
which parameterises the coset $\SL(2,\RR)/\SO(2)$. In the Euclidean
theory, on the other hand, a real theta term leads to coupling
constants parameterising $\SL(2,\RR)/\SO(1,1)$.  As will be discussed in
the next section, this is the coset structure required by holography,
and so the holographic dual should be one in which the theta-term is
real, not imaginary.

In the Euclidean theory, there are then two choices of action, one
with real theta-term and one with an imaginary one.  The imaginary one
is the one that is conventionally used in topological field theory and
is naturally associated with a real Lorentzian action.  The action
with a real theta is the one that is natural from reduction from 9+1
dimensions, and is also the one that is required for holography---the
holographic dual of the $\text{IIB}^*$ theory in de Sitter space is a
Euclidean super-Yang-Mills theory with a {\it real } theta-term.  The
formulae in this paper have all been written for the imaginary case,
but it is straightforward to obtain the real case by taking $\theta
\to i \theta$ throughout.

The Lorentzian Yang-Mills theory has a classical $\SL(2,\RR)$ symmetry
broken to a discrete $\SL(2,\ZZ)$ S-duality symmetry in the quantum
theory.  The Euclidean and Euclideanised theories also have a
classical $\SL(2,\RR)$ symmetry.  This can be seen by considering the
Yang-Mills theory from a dimensional reduction from a 6 dimensional
theory, which could be 6-dimensional (1,1) supersymmetric Yang-Mills,
or the (2,0) supersymmetric tensor multiplet theory.  The Lorentzian
and Euclidean theories can be obtained by reducing from 5+1 dimensions
on a Euclidean or a Lorentzian 2-torus, respectively.  Wick rotating
5+1 dimensional (1,1) supersymmetric Yang-Mills to a Euclideanised
theory in 6 dimensions and then reducing on a 2-torus yields the
Euclideanised four-dimensional gauge theory.  In each case, there is a
reduction on a 2-torus and so there is an $\SL(2,\ZZ)$ symmetry of the
reduced theory resulting from the large diffeomorphisms on the
2-torus.  On truncating to the massless Yang-Mills sector in 4
dimensions, the $\SL(2,\ZZ)$ is enhanced to an $\SL(2,\RR)$ classical
symmetry.  In the Euclideanised or Lorentzian theories, this is then
broken to the subgroup $\SL(2,\ZZ)$ preserving the periodic
identification of the angle $\theta$.  It is natural to conjecture
that all three theories in fact have a $\SL(2,\ZZ)$ S-duality symmetry
in the full quantum theory, and moreover that the twisted theory also
has an S-duality symmetry.


\section{Topological holography}
\label{sec:holography}

The Lorentzian $\Neq4$ supersymmetric Yang--Mills theory has a dual
holographic description as IIB string theory on $\AdS_5\times S^5$
\cite{Maldacena}, with the anti-de Sitter space formulation giving a
dual description of the Yang-Mills theory at large 't~Hooft coupling.
Wick-rotating this dual pair, the Euclideanised $\Neq4$ Yang--Mills
theory has a holographic formulation on $H^5\times S^5$, the Euclidean
version of $\AdS_5\times S^5$ with the anti-de Sitter space continued
to the hyperbolic space $H^5$ \cite{Wittenholo,Hull-Timelike}.  It was
argued in \cite{Hull-Timelike} that the holographic dual of the
Euclidean $\Neq4$ supersymmetric Yang--Mills theory is the
$\text{IIB}^*$ string theory on $\dS_{5} \times H^{5}$, where $\dS_5$ is
five dimensional de Sitter space, with the Euclidean conformal group
$\SO(5,1)$ arising as the de Sitter group, and the R-symmetry group
arising from the isometries of $H^5$.  These three dualities arise
from considering D3-branes, Wick-rotated D3-branes (i.e., instantonic
D-branes) and the Euclidean E4-branes of \cite{Hull-Timelike},
respectively.  Whereas the D3-branes are timelike 4-surfaces in 9+1
dimensions with a Lorentzian super Yang-Mills world-volume theory, the
E4-branes are spacelike 4-surfaces in 9+1 dimensions with a Euclidean
super Yang-Mills world-volume theory.

The Euclidean supersymmetric Yang--Mills theory can be twisted in four
ways, the A,B and half-twisted models, and the new conformal twisting
presented here.  It is natural to ask whether these twisted models
could still have a holographic description in de Sitter space, or more
generally what the dual description of each theory should be at large
't~Hooft coupling.  One motivation is that the de Sitter formulation
could give a useful alternative way of calculating topological
invariants.  As the 5-dimensional de Sitter symmetry is associated
with the conformal symmetry in 4-dimensions, the most promising theory
to consider in this context is the conformal twisting, as this is a
conformal field theory.  In \cite{Hull-Timelike}, it was proposed that
this conformal twisting should indeed have a holographic dual, with
correlation functions in the Euclidean conformal field theory
associated with the partition function of a theory in 5-dimensional de
Sitter space.

If $J$ is some composite operator in the Yang-Mills theory, then
introducing a source term $\int \phi(x) \cdot J$ with local source $
\phi(x)$, which can be thought of as a position-dependent coupling
`constant', then $\left<\exp( -\int \phi\cdot J)\right>$ is the
generating functional for correlation functions
$\left<J(x_1)....J(x_n)\right>$.  This in turn is related, according
to the holography conjecture, to the bulk partition function subject
to boundary conditions in which the boundary values of certain bulk
fields are given by $\phi(x)$, and for many purposes it is useful to
think of the Yang-Mills theory as living on that boundary.  In the
Lorentzian case, the boundary is the boundary of $\AdS_5$, and in the
Euclideanised theory it is the boundary of $H^5$.  For the Euclidean
case, the situation is more subtle.

The scalar fields in the Euclidean Yang-Mills theory take values in
the Lorentzian space $\RR^{5,1}$ and the expectation value of the
scalars is a vector in $\RR^{5,1}$ which can be spacelike or timelike
(or null), and these correspond to different sectors of the theory.
This corresponds to whether the separation between the E4-branes that
is kept constant in the Maldacena-type limit is spacelike or timelike
(or null).  For the spacelike separation, the arguments of
\cite{Hull-Timelike} give a holographic duality between the
$\text{IIB}^*$ theory on $\dS_5\times H^5$ and a Euclidean Yang-Mills
theory that lives on the boundary of $H^5$, while for the timelike
separation, the gauge theory lives on a boundary of de Sitter space.
This means that the boundary conditions on the boundaries of $\dS_5$ or
$H^5$ are reflected in different sectors of the Euclidean conformal
field theory \cite{Hull-Timelike}, a phenomenon that occurs in other
examples of holography in which the bulk space is a product of two
factors, both of which have a boundary \cite{HullKhuri,HulldS}.

An important role is played by those operators $J$ which lie in the
superconformal current multiplet, the multiplet of the conformal
supersymmetry consisting of the the energy-momentum tensor $T_{mn}$
and its superpartners.  These couple to fields $\phi(x)$ which lie in
the $\Neq4$ conformal supergravity multiplet, consisting of a
background metric $g_{mn}$ coupling to $T_{mn}$, and its
superpartners.  In the Lorentzian case, this is the standard $\Neq4$
conformal supergravity multiplet of \cite{Bergshoeff}, and one-loop
quantum corrections induce the conformal supergravity action
\cite{LiuTseytlin, BGMR}.  The fields in the superconformal gravity
multiplet provide the boundary conditions for the bulk fields in the
5-dimensional gauged supergravity multiplet.  Euclideanising gives a
similar picture, involving the Euclideanised conformal gravity and a
Euclideanised gauged supergravity on $H^5$.

For the Euclidean case, the situation is again similar, with a
conformal supergravity multiplet in 4 Euclidean dimensions playing a
central role.  This conformal supergravity arises from dimensional
reduction. There is a conformal supergravity in 9+1 dimensions, and
reducing on a spatial 6-torus gives the $\Neq4$ conformal supergravity
in 3+1 dimensions \cite{Bergshoeff}, while reducing on 5 space and one
time dimension in a similar manner gives a $\Neq4$ conformal
supergravity in 4 Euclidean dimensions, with gauge group $\SO(5,1)$
instead of $\SO(6)$. Again, a conformal supergravity action is induced
by one-loop corrections.  In one sector of the theory, the natural
effective supergravity theory consists of fields in $\dS_5$ which are
independent of the $H^5$ coordinates, and in the other sector it
consists of fields in $H^5$ which are independent of the $\dS_5$
coordinates, The gauged supergravity theory in de Sitter space that
arises here \cite{Hull-Timelike,HulldS} has gauge group $\SO(5,1)$ and
a twisted supersymmetry, and the boundary conditions for these fields
are provided by the conformal supergravity fields on the de Sitter
boundary.  Similarly, there is an $\SO(5,1)$ gauged supergravity on
the Euclidean space $H^5$ arising from reducing the $\text{IIB}^*$
theory on $\dS_5$, and the boundary conditions for these fields are
provided by conformal supergravity fields on the boundary of $H^5$.
Both the $\text{IIB}^*$ theory and the four-dimensional Euclidean
conformal supergravity have a pair of scalar fields taking values in
$\SL(2,\RR)/\SO(1,1)$ instead of the usual $\SL(2,\RR)/\SO(2)$.  The
boundary values of these scalars give rise to the coupling constant
and theta parameter in the dual Yang-Mills theory, with the
consequence that the theta-term is real, not imaginary, in this case.
   
We now turn to the question of finding the holographic dual for the
twisted gauge theory discussed here.  The Euclidean Yang-Mills theory
has a current supermultiplet $T_{mn},R_{mA},...$ where $R_{mA}$ is the
supercurrent.  The twisting and field redefinitions discussed here
then leads to a current multiplet $T_{mn},R_{m},\tilde R_m,...$ for
the twisted theory, where $R_m,\tilde R_m$ are BRST currents. The
fields $g_{mn},\psi_{m},\tilde \psi_m,...$ coupling to this current
multiplet then define a twisted form of the superconformal gravity
multiplet, with $\psi_{m},\tilde \psi_m$ gauge fields for local BRST
transformations.  These should give the boundary conditions for a
5-dimensional gravity theory in $\dS_5$ or $H^5$, with fields
$g_{MN},\psi_{M},\tilde \psi_M,...$ and it is straightforward to read
off at least the field content and linearised theory from the
structure of the conformal gravity multiplet, as in \cite{LiuTseytlin,
  BGMR}. This then should constitute the supergravity limit of the
holographic dual of the topological conformal field theory. In
particular, it is a theory with {\it local} BRST symmetries. For any
bosonic background, and in particular for any manifold, a rigid BRST
transformation with constant parameter will be a symmetry---the
analogue of the Killing spinor conditions are trivially
satisfied---and that rigid BRST symmetry can be used to define a BRST
cohomology, giving a gravity theory which is a topological field
theory.  We will give further details of this construction elsewhere.

\newpage

\acknowledgments

JMF is a member of EDGE, Research Training Network HPRN-CT-2000-00101,
supported by The European Human Potential Programme.  PdM was
supported by a PPARC Postgraduate Studentship PPA/S/S/1999/02882.  BS
would like to acknowledge the hospitality of Dr. H Luckock and the
School of Mathematics and Statistics at the University of Sydney,
where part of this work was done.  JMF, CMH and BS would like to
acknowledge the hospitality of the Erwin Schr\"odinger Institute in
Vienna, where this work was completed.

\providecommand{\href}[2]{#2}\begingroup\raggedright\endgroup

\end{document}